\documentclass[conference]{sig-alternate}

\usepackage{xcolor}
\usepackage{balance}
\usepackage{graphicx}
\usepackage{caption}
\usepackage{amssymb,amsfonts,amsmath}
\usepackage[numbers,sort&compress]{natbib}
\usepackage{booktabs}
\usepackage{url}
\usepackage[caption=false,font=footnotesize]{subfig}

\usepackage{paralist}
\usepackage[shortlabels]{enumitem}
\setlist{align=left,leftmargin=1.778em,labelwidth=1.278em}

\newcommand{\overbar}[1]{\mkern 1.5mu\overline{\mkern-1.5mu#1\mkern-1.5mu}\mkern 1.5mu}

\newcommand{\ceil}[1]{\lceil  #1 \rceil}
\newcommand{\vol}{\mbox{vol}}
\newcommand{\Li}{\mbox{Li}}

\newcommand{\cB}{\mathcal{B}}
\newcommand{\cN}{\mathcal{N}}
\newcommand{\sN}{\textsf{N}}
\newcommand{\cW}{\mathcal{W}}

\newcommand{\oW}{\overbar{W}}
\newcommand{\ocW}{\overbar{\mathcal{W}}}
\newcommand{\ocB}{\overbar{\mathcal{B}}}

\newcommand{\av}[1]{\langle #1 \rangle}
\newcommand{\Av}[1]{\left\langle #1 \right\rangle}
\newcommand{\eps}{\varepsilon}
\newcommand{\ones}{\underline{1}}

\usepackage{blindtext}

\usepackage{etoolbox}
\makeatletter
\patchcmd{\maketitle}{\@copyrightspace}{}{}{}
\makeatother

\newcommand{\eat}[1]{}
\newtheorem{thm}{Theorem}[section]
\newtheorem{definition}[thm]{Definition}

\begin{document}

\title{Online Myopic Network Covering}

\author{
K.\ Avrachenkov$^1$, P.\ Basu$^2$, G.\ Neglia$^1$, B.\ Ribeiro$^3$\thanks{\scriptsize Corresponding author: {ribeiro@cs.umass.edu}}, and D.\ Towsley$^3$  \\
~ \\
UMass Technical Report UM-CS-2012-034\\
\\
\centering \normalsize
       \begin{tabular}[t]{ccc}
       $^1$ INRIA & $^2$Raytheon BBN Technologies  & $^3$ Department of Computer Science       \\
       06902 Sophia Antipolis & Cambridge, MA& University of Massachusetts Amherst   \\
       France & pbasu@bbn.com & 140 Governors Drive     \\
       \{konstantin.avratchenkov, &  & Amherst, MA 01003   \\
       giovanni.neglia\}@inria.fr &  & \{ribeiro,towsley\}@cs.umass.edu
        \end{tabular}
}

\maketitle

\begin{abstract}
Efficient marketing or awareness-raising campaigns seek to recruit $n$ influential individuals -- where $n$ is the campaign budget -- that are able to cover a large target audience through their social connections. So far most of the related literature on maximizing this network cover assumes that the social network topology is known. Even in such a case the optimal solution is NP-hard.  In practice, however, the network topology is generally unknown and needs to be discovered on-the-fly.
In this work we consider an unknown topology where recruited individuals disclose their social connections (a feature known as {\em one-hop lookahead}). The goal of this work is to provide an efficient greedy online algorithm that recruits individuals as to maximize the size of target audience covered by the campaign.

We propose a new greedy online algorithm, Maximum Expected $d$-Excess Degree (MEED), and provide, to the best of our knowledge, the first detailed theoretical analysis of the cover size of a variety of well known network sampling algorithms on finite networks.
Our proposed algorithm greedily maximizes the expected size of the cover.
For a class of random power law networks we show that MEED simplifies into a straightforward procedure, which we denote MOD (Maximum Observed Degree).
We substantiate our analytical results with extensive simulations and show that MOD significantly outperforms all analyzed myopic algorithms. We note that performance may be further improved if the node degree distribution is known or can be estimated online during the campaign.

\end{abstract}



\vspace{-0.05in}
\section{Introduction}
This paper addresses the need to efficiently select $n$ individuals in a network such that they cover, through their neighbors, the largest possible fraction of the network.
Online social networks have generated much attention as a breeding ground for new forms of social studies,
social mobilization, and online campaigns.
Recruiting individuals from a population -- for instance, recruiting volunteers to get their friends to vote in an election --
is no easy task.
The recruitment of each individual comes at a cost in time, money, and social capital; and the total budget is often small with respect to the total population.
Moreover,  recruitment  is frequently targeted towards a subpopulation -- say, individuals that will likely vote for a given candidate -- that may be a relatively small fraction of the whole population.
Most works on network cover, e.g.~\cite{MaiyaKDD,Guha1998,Garey1990}, either consider the social network topology to be known in advance or assume the capability of a {\em two-hop lookahead} (where the identity of all nodes within a two-hop neighborhood of a recruited node are known), which is often not the case in the wild.

In this work we look at the cover problem when the network topology is unknown.
Following previous literature, we assume that any individual in the network can be recruited -- but in our case recruitments
mostly happen through friends recruiting friends. This  link-tracing technique has been long used by social scientists to sample  hard-to-reach subpopulations~\cite{Krista,RDS,Salganik}.
The homophily often present in social networks -- the tendency for similar individuals to be friends~\cite{MSC} -- enables the likely effective recruitment of individuals that are either in the target subpopulation or know many unrecruited individuals in the target subpopulation.
This is achieved simply by asking each recruited individual to refer other target individuals.

The recent 2012 U.S.\ presidential election presents a real-life example of an application of link-tracing recruitments to maximize the network cover of a target subpopulation. A candidate's Facebook app asked its subscribers to send get-out-to-vote reminders to their like-minded friends in swing states~\cite{FB}. 
Thus, the effectiveness of a subscriber is measured by the amount of its friends that live in swing states.
Moreover, these messages also raised awareness of the app itself, allowing it to spread through the target subpopulation of interest (see also Bond et al.~\cite{VoteNature} for a description of a get-out-to-vote Facebook app experiment in the 2010 U.S.\ elections).


\subsubsection*{Problem Formulation}
We formulate the target subpopulation cover problem
as a maximum {\em connected} cover (MCC) problem on an unknown connected graph $G=(V,E)$ (we also refer to $G$ as a network), where $V$ is the set of target individuals and $E$ the set of individuals' mutual connections.
We assume all graph parameters are unknown.
Our analysis can be easily extended to a disconnected network by considering each connected component separately.

Our main goal is to design efficient online greedy algorithms to solve the following problem on $G$: let $n$ be a given campaign budget;
we want to determine a group of $n$ individuals to be recruited in order to maximize the size of the covered subset, i.e.~of the set including the recruited nodes and their neighbors. Our only initially available information is a single node sampled from the population. Later we can acquire the neighborhood of any recruited node. It follows that the recruited nodes form a connected subgraph. 

More formally, let $\cB(t)$ be the set of $t$ known
target individuals after $t$ recruitments (also denoted the $t$-th step).
Let $\cB(0)$ be the set containing the initially known individual\footnote{\scriptsize%
	The analysis can be extended to consider many initially known individuals.
	Note that the task of finding the initial set of nodes in the target subpopulation 
	is a problem on its own~\cite{Adamic,LCCLS02}.
}
Let $\cN(\cB(t))$ be a function that returns the set of unrecruited neighbors of $\cB(t) \subseteq V$; Fig.~\ref{f:BGW} illustrates $\cB(t)$ and $\cN(\cB(t))$.
The online algorithm proceeds as follows: at step $t$, $0 <  t \leq n$, the algorithm recruits node $v \in \cN(\cB(t-1))$
and performs the update $\cB(t) = \cB(t-1) \cup \{v\}$.
The ability to obtaining the identities of the neighbors of recruited nodes, $\cN(\cB(t))$, $\forall t$, is known
as {\em one-hop lookahead} in the graph sampling literature~\cite{Adamic,MST06}.
The objective of the online algorithm is to try to maximize the size of the network cover set  $\ocW(t) = \cB(t) \cup \cN(\cB(t))$, for $t=1,\ldots, n$,
without having a priori access to topology information.
We refer to the problem of online covering an unknown network in the presence of one-hop lookahead  as the
 {\em online myopic network covering} problem.

\subsubsection*{Contributions}
We make the following contributions:\\

\vspace{-0.05in}
\noindent
{\bf (1)} We thoroughly evaluate -- analytically and through extensive simulations on social network datasets%
\footnote{ \scriptsize%
See Sec.~\ref{sec:related} for some limited analysis of other ``non-social'' networks.
}
 -- the performance of several known network sampling algorithms.
{\bf (1.1)} We  investigate the cover sizes of Breadth-First Search (BFS) and Depth-First Search (DFS).
We observe a consistent large variance in the cover sizes found by BFS and that BFS tends to underperform in comparison to a greedy oracle scheme that recruits at every step the node with the largest number of uncovered neighbors.
We partially blame network homophily for the lack of performance from BFS.
DFS, which at first sight should improve upon BFS in circumventing the above homophily problem, performs even worse.
Using random networks, we show why DFS finds a small cover after $t \ll N$ recruited nodes. 
{\bf (1.2)} A Random Walk (RW), more precisely  RW {\em without replacement} (RWnr), where nodes revisited by the walker are not counted towards the recruitment budget, is shown to consistently outperform (sometimes significantly) BFS in our simulations.
{\bf (1.3)} We propose a new online algorithm inspired by the Susceptible-Infected (SI) epidemic model but observe that RWnr is consistently more efficient than SI.
\\

\vspace{-0.05in}
\noindent
{\bf (2)} Our work is, to the best of our knowledge, the first to provide an analytical characterization of the sizes of $\ocW(t)$ (the cover) as a function of $t$ (recruited nodes), for RWnr
and the SI epidemic algorithms on {\em finite networks}. As recently acknowledged in~\cite{FPS12}, this was a challenging open problem.
Moreover, we establish an interesting connection between cover through RWs and the coupon subset collection problem.
We validate our theoretical results through simulations.
\\


\vspace{-0.05in}
\noindent
{\bf (3)} We propose a new online algorithm (MEED, Maximum Expected $d$-Excess Degree) that greedily maximizes the expected size of the cover.
For a broad class of power law networks, MEED simplifies into a straight forward heuristic, which we denote Maximum Observed Degree (MOD).
We substantiate our analytical results with simulations.
Extensive simulations on a variety of social network datasets show that MOD consistently outperforms (sometimes significantly) all other analyzed algorithms. Performance can be further improved if  the node degree distribution is known or can be estimated online during the campaign.

\subsection*{Outline}
The reminder of this work is organized as follows.
Sec.~\ref{sec:notation} presents the notation and background used throughout this work.
Sec.~\ref{sec:opt} discusses optimal solutions and approximations in connection to the connected minimum dominating set.
Sec.~\ref{sec:datasets} presents the datasets used in this work and our simulation setup.
Sec.~\ref{sec:BFS_DFS} provides an analysis of the effectiveness of Breadth-First-Search (BFS) and Depth-First-Search (DFS).
Sec.~\ref{sec:RW} provides a deep analysis of the effectiveness of two types of random walks and compare them to BFS. 
Sec.~\ref{sec:SI} proposes a sampling algorithm inspired by Susceptible-Infected (SI) epidemic models. 
We also provide an analytical solution describing the cover size of SI as a function of $t$.
An important feature of our analysis is our ability to model finite graphs, which is key to understanding the effectiveness of large campaigns in respect to the size of the target population.
Sec.~\ref{sec:excess} proposes MEED and MOD as a simple approximation of MEED.
Sec.~\ref{sec:excess}  also provides theoretical and simulation results, the latter comparing MOD against the other algorithms.
And, finally, Sec.~\ref{sec:related} summarizes our contributions and reviews the related work.


\section{Notation \& Background}
\label{sec:notation}

\begin{figure}[h!]
\centering
\vspace{-0.05in}
\includegraphics[scale=0.2]{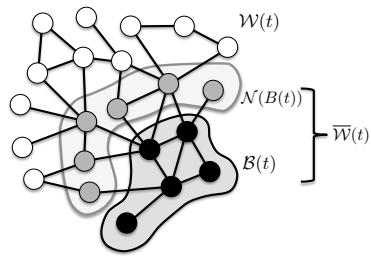}
\vspace{-0.25in}
\caption{Network sampling evolving sets.\label{f:BGW}\label{f:notation}}
\vspace{-0.1in}
\end{figure}

We consider an unknown connected network $G=(V,E)$ with $N=|V|$ nodes, $M=|E|$ edges, and degree distribution $\{p_k\}_{k=1,\dots N-1}$.
We assume all graph parameters are unknown to us.
Denote $\cN_a(v)$ the  set of neighbors of node $v \in V$, irrespective of their recruitment status, and $k_v = |\cN_a(v)|$ is the degree of $v$.
For each step $t = 1,\ldots, n$, where $n \in \{1,\ldots,N-1\}$ is the campaign budget, we classify the nodes in $V$ into three disjoint sets.
The set $\cB(t)$ denotes the recruited nodes at step $t$; these are the black nodes in Fig.~\ref{f:notation}.
Unrecruited neighbors of recruited nodes are denoted {\em observed nodes} and form the set $\cN(\cB(t))  = \cup_{v \in \cB(t)} \cN_a(v) - \cB(t)$ (gray nodes in Fig.~\ref{f:notation}).
We say a node $v \in V$ is {\em covered} at step $t$ if $v \in \ocW(t)$, where $\cW(t)=V - \cB(t) \cup \cN(\cB(t))$ is the set of all \emph{uncovered} nodes (white nodes in Fig.~\ref{f:notation}) and $\ocW(t) = V - \cW(t)$ its complement.
Note that at time $t$ we are unaware of the existence of nodes in $\cW(t)$.

The sizes of the three sets $\cB(t)$, $\cN(\cB(t))$, and $\cW(t)$ are denoted $B(t)$, $\sN(\cB(t))$ and $W(t)$, respectively.
Clearly $B(t)+\sN(\cB(t))+W(t)=N$ at any step $1 \leq t \leq n$.
Finally, for the sake of simplicity, we allow a slight abuse of notation, denoting by $\av{.}$ both the empirical mean and the expected value. The exact interpretation of $\av{x}$ will then depend on the nature of the quantity $x$.
We use the convention that $\av{k}$ denotes the average degree.
Table~\ref{t:notation}  summarizes the notation used throughout the paper.

Our analysis makes extensive use of the configuration random graph model~\cite{NewmanConf}. This is a defined as a uniform probability distribution over the ensemble of the graphs where nodes have a given degree distribution $\{p_k\}_{k=1,\dots N-1}$.
A configuration model sample can be generated as follows. The degree $k_i$ is attributed to each node $i$ according to the selected degree distribution. Each vertex $i$ can then be thought of having $k_i$ stubs attached to it that are the ends of edges-to-be. By connecting randomly selected stubs' pairs the graph sample is generated.

The configuration model is widely used in the complex network literature~\cite{Newman,Vespignani} also for the simplicity of the analysis. Moreover, as we will soon see, the formulas we derive to predict the value of $\sN(\cB(t))$ 
considering the configuration model match the results of our simulation for actual topologies remarkably well.

\begin{table}
\begin{tabular}{ll}
\toprule
Variable & Description \\
\midrule
$N$ & no.\ of nodes \\
$M$ & no.\ of edges \\
$\cN_a(v)$ & set of neighbors of $v \in  V$\\
$\av{x}$ & average value of quantity $x$\\
$p_k$ & fraction of nodes with degree $k$ \\
$\cB(t)$, ($B(t)$) & set (number) of sampled nodes at step $t$\\
$\cN(A)$, ($\sN(A)$) & set (no.) of unrecruited neighs of $A \subseteq V$\\
$\cW(t)$, ($W(t)$) & set (no.) of uncovered nodes at step $t$\\
\bottomrule
\end{tabular}
\vspace{-0.17in}
\caption{Notation Table\label{t:notation}}
\vspace{-0.2in}
\end{table}

%
\section{Network Covers,  Oracles, \\ \& Approximate Solutions}
\label{sec:opt}
The problem we study is closely related to the well-studied  Maximum coverage~\cite{nemhauser78} and Minimum Connected Dominating Set (MCDS) problems.
The maximum coverage problem can be described in our setting as selecting at most $n$ nodes such that the union of the nodes they cover has maximal size.
The maximum coverage problem is NP-hard, and cannot be approximated within $1 - 1/e + o(1)$, where $e$ is the Euler constant. A simple submodular function greedy algorithm, however, is able to find a $1 - 1/e$ approximation~\cite{nemhauser78}.
 Our problem setting, however, requires the recruited nodes to be connected to each other.
In the connected setting, above mentioned greedy algorithm is similar to a greedy algorithm used to solve the MCDS, described as follows.

Given a graph $G=(V,E)$ with $N$ nodes, $DS \subseteq V$ is a dominating set if $\forall_{v\in V}: v \in DS \textrm{ or } v \in V - DS$. Thus, if all nodes in $DS$ are recruited as {\em dominators}, it may be possible to reach all nodes in the network through these {\em dominators}. 
The MDS problem is to find the set $DS$ with the minimum cardinality. MCDS imposes an additional restriction that the subgraph induced by the vertices in $DS$ has to be connected.
\eat{Since influence spreads through the network, the most natural way for an influential person to spread his/her influence is through his/her direct contacts (or neighbors). Under this assumption, it is not enough to compute the MDS of $G$ since the dominators may not know each other; instead, one has to aim to compute the Minimum {\em Connected} Dominating Set (MCDS) of $G$, which is essentially an MDS with an additional restriction that the subgraph induced by the vertices in $DS$ mentioned above has to be connected.}

Our goal is not to cover {\em all} of $G$; instead, we seek to cover as much of $G$ as possible with recruitment budget $n$. 
However, since MCDS is very closely related to our problem, key results and techniques from the MCDS literature can provide crucial insights into the role of lookahead in network coverage especially about worst-case performance guarantees when compared to the optimal solution. In a situation where complete network knowledge is available\footnote{\scriptsize%
This could arise in ``intelligence gathering" applications where analysts have pieced together the topology of an adversary network and now want to recruit the best (connected) set of influencers that will {\em cover} it.}, solving MCDS is NP-hard~\cite{Garey1990}. However, there exist well-known linear approximation preserving reductions (L-reductions~\cite{Kann1992}) from the {\sc SetCover} problem to MCDS~\cite{Guha1998} that yield a guaranteed approximation factor of $O(\ln{n})$.

\vspace{-0.05in}
\begin{definition}[Observed degree] The number of recruited neighbors of a node.
\end{definition}
\vspace{-0.05in}

\vspace{-0.1in}
\begin{definition}[$d$-excess degree] A node with degree $k$ and observed degree $d \leq k$ has excess degree $k - d$.
\end{definition}
\vspace{-0.05in}

If there is limited ``lookahead", say, {\em two-hop} information of the neighborhood of each recruited node, the natural algorithm is to greedily recruit nodes that have the maximum number of uncovered neighbors, i.e., with the maximum {\em excess degree}. Guha and Khuller~\cite{Guha1998} implemented this greedy algorithm by building {\em growing} a tree $T$ in an online fashion, starting from a single node. Initially all nodes are unrecruited (white). At each step, a vertex $v \in T$ with the largest excess degree is recruited (colored black) and edges are added to $T$ which exist between $v$ and all its neighbors that are not in $T$ (these unrecruited neighbors are colored gray). The algorithm stops when all nodes are colored either gray or black, and the connected dominating set (CDS) is the set of non-leaf nodes in $T$. They showed that the above algorithm has a guaranteed approximation ratio of $O(\Delta)$, where $\Delta$ is the maximum degree of the network. We refer to the aforementioned algorithm as ``Oracle" as it requires two-hop lookahead in order to compute the excess degree of nodes in $\cN(\cB(t))$), a capability often missing in real online social networks. 
An example of an implementation of Guha and Khuller's Oracle can be found in Maiya and Berger-Wolf~\cite{MaiyaKDD} (denoted {\em Expansion Sampling} in their work).

Interestingly, Guha and Khuller also showed that the approximation factor can be significantly improved when {\em three-hop lookahead} is exploited in a modified greedy step: recruit a {\em pair} of adjacent vertices (i.e. mark them black) and compare the yield in the number of gray nodes acquired in the neighborhood of this pair; at each step greedily select a pair of vertices or a single vertex that maximizes this yield. This modified greedy step surprisingly yields an approximation ratio of $O(\ln{\Delta})$ instead of $O(\Delta)$. This additional lookahead, however, is not be available in several practical settings that we are interested in studying in this paper.

In practice, with {\em one-hop lookahead}, only {\em observed degree} information is available at nodes in $\cB(t)$. 
This results in our MEED algorithm, which uses Guha and Khuller's Oracle approach using the expected excess degree in place of the true excess degree.
MEED, however, requires the degree distribution of the network as side information.
In the absence of degree distribution information, we show that for some random power law networks, a natural myopic online greedy algorithm of recruiting the node with the maximum observed degree approximates MEED -- this is our MOD algorithm and also Maiya and Berger-Wolf's SEC scheme~\cite{MaiyaKDD}.
\eat{We conjecture that even with such online myopic constraints, the competitive ratio is still $O(\Delta)$.}
Expected value analysis as well as simulations in Sec.~\ref{sec:excess} show that MOD is a good heuristic when operating on realistic social network such as those obeying a power law degree distribution.

We note that due the online nature of the different algorithms presented above, it is easy to apply the maximum budget criteria of MCNC and stop whenever the allocated budget $n$ has been consumed. While the theoretical approximation guarantees may not strictly apply (unless $n=N$), we believe that in practice they still hold for the networks studied in this paper.

\section{Datasets \& Simulation Setup}
\label{sec:datasets}

We use the Enron email dataset as the running example throughout this work.
The Enron email dataset contains data from a subpopulation of about 150 users, mostly senior management of Enron.
This data was made public by the Federal Energy Regulatory Commission during its investigation of Enron.
In total there are 36,692 nodes (unique email users) with average degree of $10.02$ and average clustering coefficient of $0.5$.
The high clustering coefficient suggest great homophily in this network.
The email corpus description can be found in Klimmt and Yang~\cite{Enron}.
The version of the email graph we use can be downloaded from the Stanford's SNAP repository~\cite{Snap}.

We also make use of other social network datasets -- all of them, except Flickr, are available online at SNAP~\cite{Snap}.
We now describe our datasets making use of $N$, $\av{k}$, and $c$ to denote the number of nodes, the average degree, and the clustering coefficient, respectively:
Epinions ($N= 75,\!877$, $\av{k}= 10.7$, $c=0.26$) and Slashdot ($N= 82,\!168$, $\av{k}= 0.1$, $c=0.23$) online social networks, 
Wiki-talk ($N= 2,\!394,\!385$, $\av{k}=3.9$, $c=0.2$) Wikipedia user-to-user discussion graph, EmailEU ($N=265,\!214$, $\av{k}=2.8$, $c=0.28$) the network email communication between users of a large European research institution, Youtube ($N=1,\!134,\!890$, $\av{k}=5.3$, $c=0.17$) friendship network of youtube.com users, and finally,  Flickr dataset, a snapshot of an online photosharing network with $N=1,\!715,\!255$ nodes and $\av{k} = 12.2$, collected in Mislove et al.~\cite{Mislove}. 

We also contrast our social network results with the results on three non-social networks.
These networks can also be found online at SNAP~\cite{Snap}.
Gnutella ($N=62,\!561$, $\av{k}=4.7$, $c=0.01$) a collection of merged P2P client snapshots collected in Ripeanu et al.~\cite{Gnutella}, 
HepTh ($N=27,\!770$, $\av{k}=25.4$, $c=0.3$) a paper citation graph, and Amazon ($N=334,\!863$, $\av{k}=5.5$, $c=0.4$) the network of co-purchased products on the amazon.com website. 

\vspace{-0.05in}
\paragraph{Simulation setup}
Unless stated otherwise our metrics consist of averages over $1,\!000$ simulation runs.
We use colored shadows in our plots to show the value of standard deviation plotted around the average.
The shadow serves two proposes.
First, its vertical width multiplied by $1.96/\sqrt{1000}$ gives approximately the $95\%$ confidence intervals of our averages.
Second, its value measures the variability between independent runs, by which we compare how consistently good (or bad) an algorithm performs.
In our simulations $\cB(0)$ includes a single node recruited uniformly at random from $V$.
The order in which neighbors of a node appear on its list of neighbors is randomized from run to run to avoid arbitrary biases that may arise from the choice of node IDs in the dataset.

\pagebreak


%
\vspace{-0.05in}
\section{BFS \& DFS Algorithms}\label{sec:BFS_DFS}
We begin our study by comparing the performance of two different approaches derived
from two basic graph traversal algorithms: {\bf Breadth-First Search (BFS)} and {\bf Depth-First Search (DFS)}.
BFS is chosen because it is widely used in network sampling~\cite{Mislove,OFrank,KurantJSAC11,Najork01}.
In these algorithms nodes in the $\cN(\cB(t))$ are recruited according to the time they were first observed (a node is observed when one of its neighbors is recruited).
If we consider that nodes are put in a queue when they are first observed and then removed when they are recruited, then 
BFS employs a First In First Out discipline for the queue, recruiting the first observed node in $\cN(\cB(t-1))$, while DFS employs a Last In First Out discipline, recruiting the last observed node in $\cN(\cB(t-1))$.
At each step a new, previously unrecruited node, is recruited such that at step $t=N-1$ all nodes are recruited, 
i.e., $\sN(\cB(N-1)) = 0$.

\begin{figure*}[p]
\vspace{-0.3in}
\centering
\subfloat[][ \label{f:BFS_DFS}]{
\includegraphics[width=3in,height=3in]{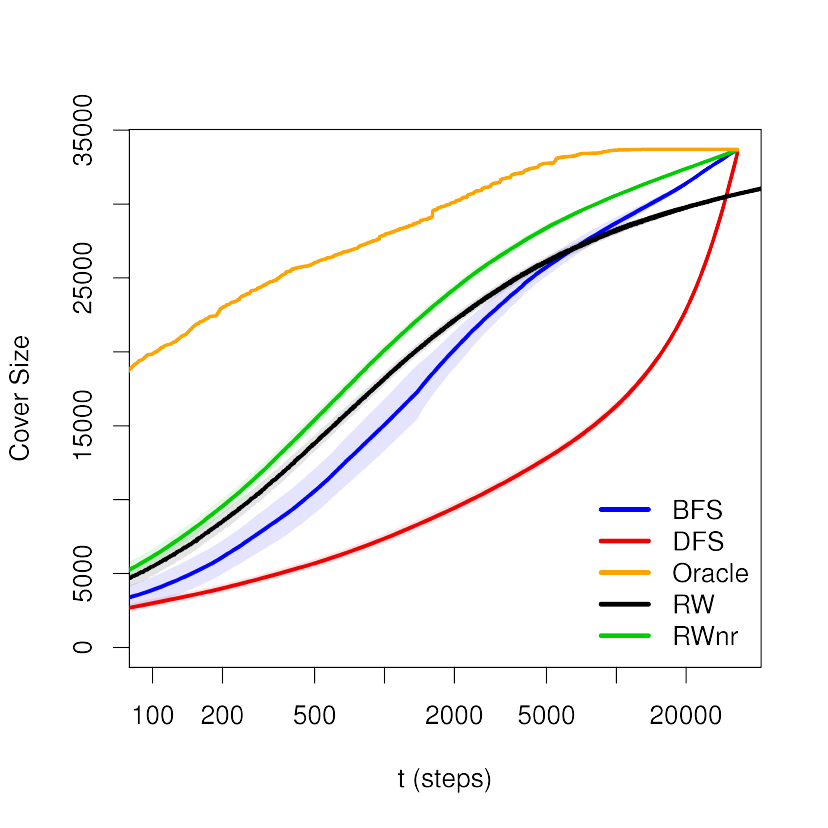}
}
\subfloat[][\label{f:GSIMOD}]{
\includegraphics[width=3in,height=3in]{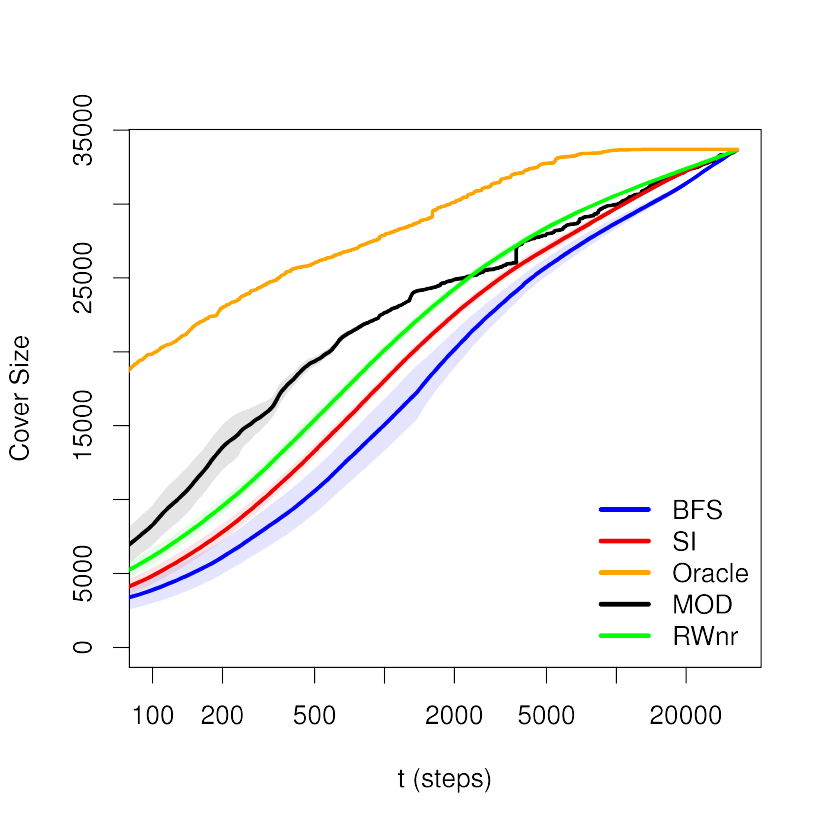}
}
\caption{\small {\bf (Enron Network)} Empirical average cover size $\av{\oW(t)}$ as a function of $t \in \{1,N-1\}$. Fig.~\ref{f:BFS_DFS} compares Oracle, RW, RWnr, BFS, DFS and Fig.~\ref{f:GSIMOD} compares Oracle, RWnr, SI, BFS, MOD. Shadows show double the standard deviation of $1,\!000$ simulations; $x$-axis in log-scale.\label{f:Enron}}
\vspace{-0.1in}
\end{figure*}

Fig.~\ref{f:BFS_DFS} shows the average cover size $\av{\oW(t)}$ of BFS and DFS as a function of $t$ on the Enron email network (recall that we average over $1,\!000$ simulation runs).
We find similar results on all of our social network datasets, see Figs.~\ref{f:social}a-f.
The simulations show while both BFS and DFS achieve the full coverage for $t \approx N$, BFS significantly outperforms DFS for all other values of $t$.
To understand this difference, we qualitatively analyze the step in which a given node $v \in V$ with degree $k_v$ is recruited.

Because both BFS and DFS follow edges to recruit nodes, the probability that $v$ is first observed in $\cN(\cB(t))$ at step $t$ is approximately $(\gamma_v/N) (1 - \gamma_v/N)^{t-1}$, where $\gamma_v = k_v/\av{k}$ (this simple formula should be a good approximation in a configuration model where nodes are recruited independently by both algorithms; it also assumes $t \ll N$).
Thus, large degree nodes tend to be observed earlier in the process than small degree nodes.
As a FIFO policy recruits the earliest observed nodes from $\cN(\cB(t))$, BFS tends to recruit large degree nodes first,
on the other hand, a LIFO policy recruits the latest observed nodes from $\cN(\cB(t))$, hence {\bf DFS tends to recruit small degree nodes first.}
This DFS result contradicts previous results in the literature~\cite{MaiyaKDD}, which we revisit in Sec.~\ref{sec:related}.

Note that Fig.~\ref{f:BFS_DFS} shows larger standard deviations for BFS than for DFS (although in some social networks the relative difference may be small).
This is because the cover size of a non-neglegible fraction of the BFS runs deviates from the average.
This instability is due to the strong dependence of the BFS cover size on the initial node $\cB(0) = \{i\}$.
As BFS explores the network in ``waves'' (expanding rings from $i$), the initial node selection may significantly impact BFS's cover size.
Moreover, {\bf we expect BFS to perform poorly on networks with a large degree of homophily} (as seen at the end of Sec.~\ref{sec:RW} in a simple regular lattice example).
Homophily is the tendency of individuals to connect to similar individuals~\cite{MSC}, thus creating patches of clustered nodes in the network.
This means that if $v \in V$ is connected to $u \in V$ and $z \in V$, then $u$ and $z$ are more likely to be connected than random chance would allow.
In addition, if $v$ is not connected to $w \in V$ then $u$ and $z$ are more likely than random not to  be connected to $w$.
In such scenario it pays not to recruit both $u$ and $z$ together, as their neighbors significantly overlap with higher probability than random chance alone would allow.

DFS clearly avoids the above homophily problem by traversing the graph in depth first order. To increase the cover set size we only need to modify the LIFO recruitment policy without resorting to BFS's FIFO policy.
In what follows (Sec.~\ref{sec:RW}) we explore the use of Random Walks (RWs).
As we see in the next section, a RW share commonalities with DFS in that it also traverses the graph from the last recruited node (however, a RW may try to recruit a node more than once).
But, different from DFS, it allows recruitment of observed nodes in $\cN(\cB(t))$ irrespective of when the node was observed.
One drawback of RWs -- that fortunately can be easily mitigated via caching -- is the possibility of recruiting already recruited nodes.

%

%
%

%
\section{RW Algorithms}
\label{sec:RW}
Now let us analyze the cover size of {\bf Random Walks (RWs) with one-hop lookahead}.
Increased attention has been paid to random walks as a tool for network sampling~\cite{RT10,Krista,Salganik,Kurant11,KurantJSAC2_11} mostly due to its good statistical properties.
In the RW algorithm $\cN(\cB(t))$ is still the set of all observed unrecruited nodes at time $t$.
However, in a RW the node to be recruited at step $t+1$ is a random neighbor of the node recruited at step $t$, regardless of the time that  the node was observed or even if it was already recruited.
We begin our analysis assuming that a node that is recruited again at step $t$ needs to be paid (that is, time advances even if no new recruitments were performed).
We refer to this traditional RW algorithm as {\bf RW with replacement (RW)}, in which nodes already recruited can be recruited again.
At the end of this section we extend our analysis to the case of {\bf RW without replacement (RWnr)}
where already recruited nodes are ``cached'' so that recruiting nodes from $\cB(t)$ does not count towards the recruitment budget.

The cover size of RWs with one-hop lookahead has been the subject of previous work~\cite{MST06}.
However, we feel that one needs to exercise caution when interpreting the results in Mihail et al.~\cite{MST06}.
Mihail et al.\ shows that a RW with one-hop lookahead finds the majority of nodes in sublinear time in an infinite configuration model with
heavy tailed power law degree distribution.
As our approach demonstrates below, covering finite networks is patently different from covering infinite networks.
In particular, we show that for any given finite sized network, the discovery rate is never superlinear (this linear growth rate, however, can be large).
Our model also allows us to predict with high accuracy the expected number
of covered nodes as a function of $0 < t \leq n$.

Let us first analyze the performance of RW.
In RW, the expected cover size at step $t$  is
\begin{equation}\label{eq:EWkRW}
\begin{aligned}
\av{\oW(t)}  & =  N - \sum_{\forall v \in V} P[\mbox{node $v$ is uncovered}] \\
& = N - {\bf q} \sum_{\forall v \in V} \ {}_{\cN_a(v)} {\bf P}^t \ones,
\end{aligned}
\end{equation}
where ${\bf q}$ is a vector with the initial distribution of the random walk,
$\ones$ is a column vector of ones,  and ${}_{\cN_a(v)}{\bf P}$ is a taboo transition
probability matrix defined by
$$
[{}_{{\cN_a}(v)}{\bf P}]_{ij}=
\left\{ \begin{array}{ll}
p_{ij}, & \mbox{if} \ i,j \not \in {\cN}(v),\\
0, & \mbox{otherwise,}
\end{array} \right.
$$
with ${\cN_a}(v)$ denoting the neighborhood set of node $v$, including node $v$.

The above formula (\ref{eq:EWkRW}) requires complete topology knowledge
and does not allow simple analytical solution.
However, consider the following approximation to a RW.
Nodes are recruited with replacement in i.i.d.\ fashion according
to the stationary distribution of the random walk. Then, the expected cover size at step $t$ would be given by
\begin{equation}\label{eq:EWkRWsteady}
\begin{aligned}
\av{\oW(t)} & = N - \sum_{\forall v \in V} P[\mbox{node $v$ is uncovered}] \\
   & = N - \sum_{\forall v \in V} (1-\alpha_v)^t,
\end{aligned}
\end{equation}
\vspace{-0.1in}
where
\vspace{-0.1in}
$$
\alpha_v = \frac{1}{2M} \left(k_v + \sum_{j \in \cN_a(v)} k_j\right).
$$
The above can be interpreted as a particular case of the coupon subset collection problem~\cite{AR01,M77}.
Each step $t$, $t=1,\ldots,N-1$,  we draw a subset of ``coupons'', a subset of newly observed nodes in our terminology.
We can observe a node either by sampling it directly (this corresponds to the term $k_v/(2M)$) or
by sampling one of its neighbors (this corresponds to the term $\sum_{j \in \cN_a(v)} k_j/(2M)$).
The value of $k_v + \sum_{j \in \cN_a(v)} k_j$ is known as the second neighbor degree~\cite{Adamic}.
In Appendix~\ref{appx:RW} we use matrix perturbation theory to show that~\eqref{eq:EWkRWsteady}
well approximates~\eqref{eq:EWkRW} for fast mixing RWs (see~\cite{ART10,RT10} for fast mixing RW techniques).

\begin{figure}[h!]
\centering
\vspace{-0.2in}
\includegraphics[width=2.8in,height=2.8in]{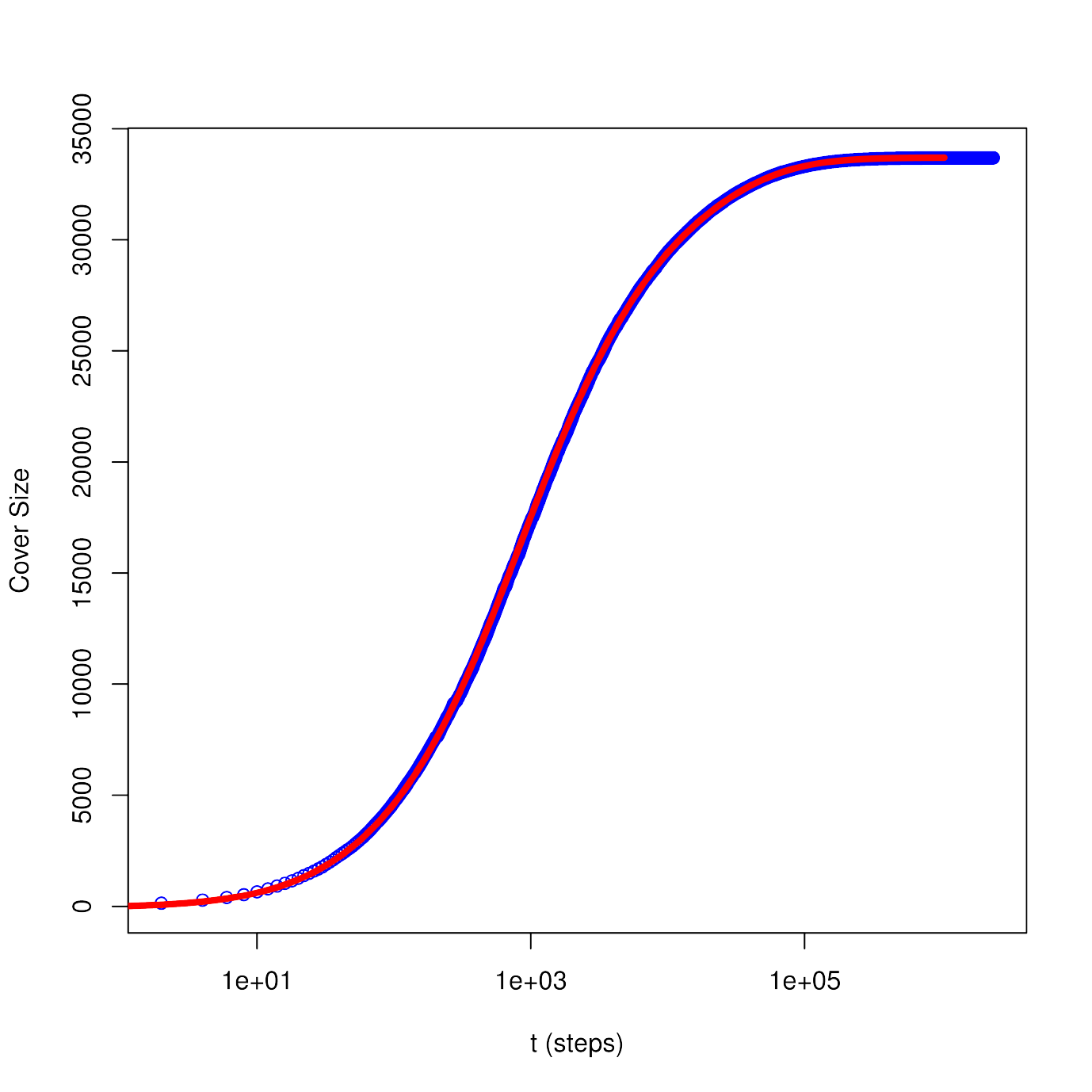}
\vspace{-0.1in}
\caption{\small%
{\bf (Enron email)} Theoretical RW cover (red line) against simulations (blue circles) on Enron email network.
Plots in semi-log scale.\label{f:Enron-modeCC}}
\end{figure}

Applying the Taylor series expansion and the fact that $2M=\av{k}N$,
we can write -- for small $t$  -- the following approximation
\vspace{-0.1in}
$$
\av{\oW(t)} \approx \frac{t}{\av{k}N} \sum_{\forall v \in V} \left(k_v + \sum_{j \in \cN_a(v)} k_j\right).
$$
Next, we note that $\sum_{\forall v \in V} \sum_{j \in \cN_a(v)} k_j = \sum_{\forall v \in V} k_v^2$,\\
which yields
$$
\av{\oW(t)}  \approx  t \frac{\av{k^2}+\av{k}}{\av{k}},
$$
where $\av{k^2}$ is the empirical second moment.
%
Thus, the cover process has approximately linear large growth rate in the initial phase, if the second moment $\av{k^2}$ is large with respect to the average $\av{k}$.
Next, let us treat~\eqref{eq:EWkRWsteady} as a continuous function of $t > 0$. The second derivative with respect to $t$  is
$$
\frac{d^2}{dt^2} \av{\oW(t)}  = - \sum_{\forall v} (1-\alpha_v)^t \log^2(1-\alpha_v) < 0,
$$
which is a concave function and, consequently,
for any fixed network, the growth rate of the cover of RW with one-hop lookahead is at most linear.
Figure~\ref{f:Enron-modeCC} shows that our theoretical result in~\eqref{eq:EWkRWsteady} accurately reflects the cover in simulations over the Enron email dataset.

Next we analyze the {\bf RWnr} algorithm.
Here, when  the walker comes across an already recruited node, it does not recruit it again but rather randomly proceeds to one of its already cached neighbors.
We define the algorithm evolution such that at each step we recruit a new node (that is, ``virtual re-recruitments'' do not advance $t$).
RWnr closely follows the derivations for RW, with the only difference
that now some recruitments cannot be ``wasted'' on already recruited nodes.
The key observation here is that we need to keep track of the edges belonging to recruited nodes.
We suggest the following approximation.
Let $Z(t)$ be the number of edges not yet discovered at time $t$,
then we approximate (details can be found in Appendix~\ref{apps:RWnr})
\begin{equation}\label{eq:EWkRWnrreplace}
\av{\oW(t)} \approx N - \sum_{v \in V} \: \prod_{i=0}^{t-1}\left(1-\frac{1}{2\av{Z(i)}}
\left(k_v + \sum_{j \in \cN_a(v)} k_j\right)\right) \, ,
\end{equation}
\vspace{-0.1in}
where
\begin{equation*}\label{eq:EZkRWnrreplace}
\av{Z(t)} \approx \sum_{(u,v) \in E} \: \prod_{i=0}^{t-1} \left(1-\frac{k_u+k_v}{2\av{Z(i)}}\right).
\end{equation*}
From~\eqref{eq:EWkRWnrreplace} we note that RWnr and privileges nodes with large second neighbor degrees (just like RW does).

The plots in Fig.~\ref{f:BFS_DFS} show that RWnr significantly outperforms RW for large values of $t$.
Note that for small values of $t$, the graphs in Fig.~\ref{f:BFS_DFS} shows that the performance of RW is similar to the performance of RWnr.
This behavior is observed because for small values of $t$, $\av{Z(t)} \approx \av{k}N$, and thus
the values of $\av{W(t)}$ in eqs.~\eqref{eq:EWkRWsteady} and~\eqref{eq:EWkRWnrreplace} are similar.

The plots in Fig.~\ref{f:BFS_DFS} also show that RWnr significantly outperforms BFS.
These results are consistent on all other datasets (refer to the plots in Figs.~\ref{f:social}a-f)
This curious fact can be explained by observation that BFS is affected by the graph homophily.

\begin{figure*}[htp]
\centering
\vspace{-0.3in}
\subfloat[][2D Lattice]{
\includegraphics[width=3in,height=3in]{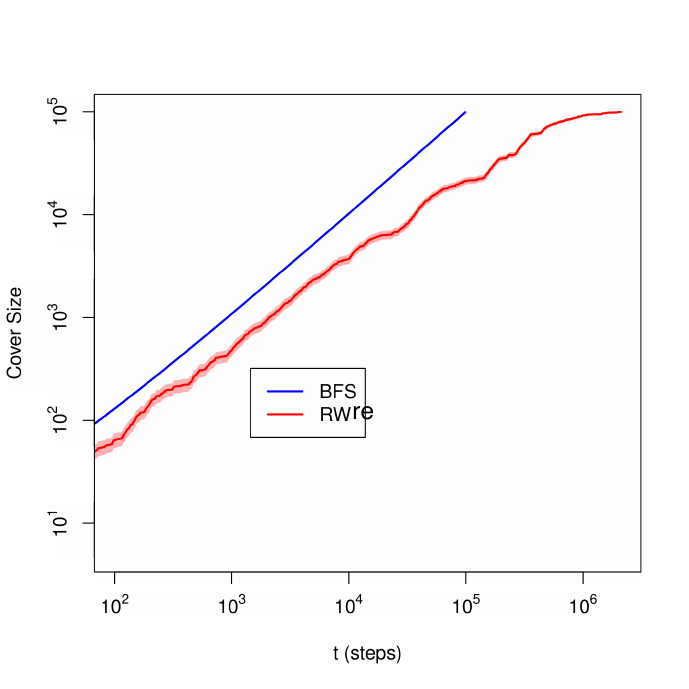}
}
\subfloat[][3D Lattice]{
\includegraphics[width=3in,height=3in]{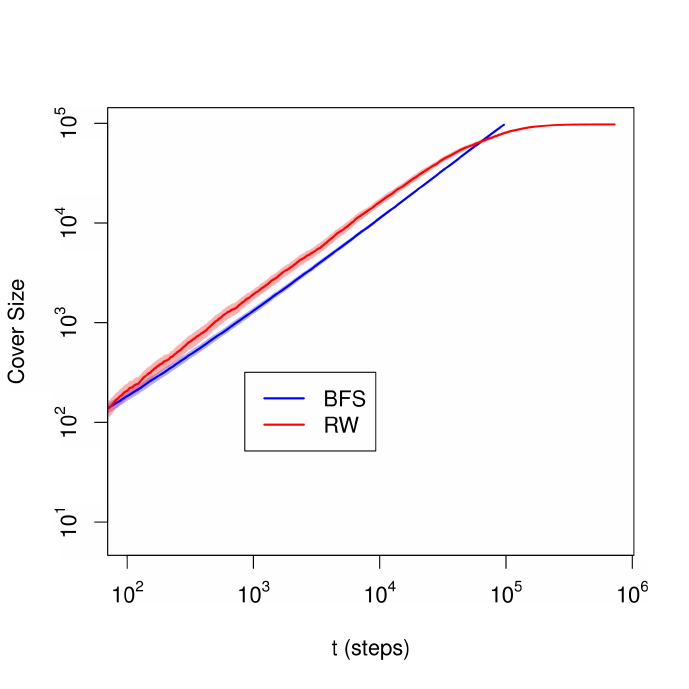}
}
\caption{\small%
$\av{\oW(t)}$ of RW against BFS on a 2D and 3D lattices.
Simulations average over $1,\!000$ runs, standard deviation shown as a colored shadow.
Plots in log-log scale.\label{f:BFSRWLattice}}
\vspace{-0.2in}
\end{figure*}

At first sight, however, it seems that both RW and RWnr should be impacted by homophily, as both have a tendency to ``over-explore'' a neighborhood, just as BFS.
However, as we see next, this depends on the network and, in social networks, we believe that this is not likely the case.
We contrast the performance of RW against BFS on two regular lattices.
We choose these graphs because on regular lattices $\av{k^2}/\av{k}$ is small,
leaving just homophily (for BFS) and the RW escape probability (the probability that a RW never revisits the same node or neighborhood again in an infinite network) as the primary factors in determining cover size for BFS and RW.
Fig.~\ref{f:BFSRWLattice} shows the performance of RW and BFS
on regular 2D and 3D lattices.
The plots show the cover size $\oW(t)$ as a function of $t$
where both lattices have approximately $N=10^5$ nodes each.
Observe that RW suffers from its tendency to return to the same nodes.
On an infinite 2D regular lattice a RW returns to the same node with probability one
while in a regular 3D lattice this probability is 0.34~\cite{W39}.
On the other hand, BFS is affected by the clustering of the lattice, such that for most recruited nodes
BFS in average covers approximately only one new node per step. A detailed analysis of these approximations can be found in Appendix~\ref{appx:grids}.

%

\section{SI Algorithm}\label{sec:SI}
In this section we consider a different method, inspired by the {\bf Susceptible-Infected (SI}) model in epidemiology:
at step $t$ recruit a node from $\cN(\cB(t))$ by randomly selecting one of the edges between $\cB(t)$ and $\cN(\cB(t))$.
Under SI node $v \in \cN(\cB(t))$ with $d(v,t)$ neighbors in $\cB(t)$
is recruited at time $t$ with probability $d(v,t)/\sum_{u \in\cN(\cB(t))} d(u,t)$.
Our analysis of RW-based cover size relied on the fact that a node is always recruited with probability proportional to the degree of that node.  This is no longer the case for SI-based cover sizes. Moreover, unlike the SI-related epidemic literature on infinite graphs~\cite{Vespignani}, we will observe from our analysis that the probability that a particular node is recruited depends on $t$, the number of steps that have been executed.

Before we delve into the analysis of SI, we first show that we cannot ignore the impact of $t$ on the degree distribution of nodes in $\ocB(t)$, which, as we see next, becomes significantly different than the degree distribution across the whole network $\{p_k\}_{k=1,\ldots}$ as $t$ gets larger.

\subsection{The effect of {\large $t$} on the degree \\ distribution of {\large $\ocB(t)$}}
\label{sec:changedist}
\begin{figure*}[t]
\centering
\vspace{-0.05in}
\includegraphics[width=4in,height=3in]{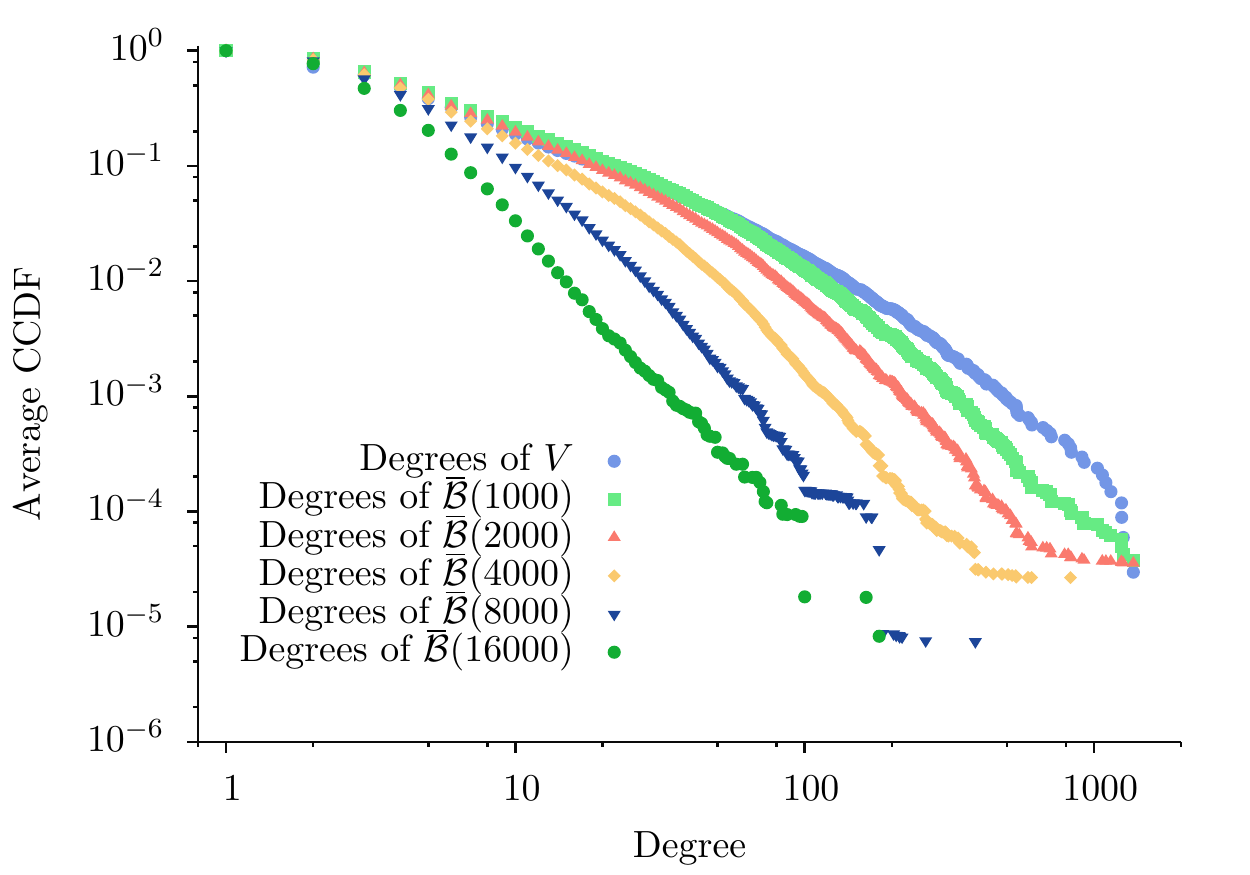}
\caption{\small%
{\bf (Enron Network)} Complementary Cumulative Distribution Function (CCDF) of $\ocB(t)$ averaged over $1,\!000$ runs, for $t\in \{1000,2000,4000,8000,16000\}$ recruited nodes in the Enron Network (Enron has approximately $N = 36,\!000$ nodes). As SI recruits nodes in the graph, the degree distribution of the remaining nodes in $\ocB(t)$ suffers a dramatic change. Plot in log-log scale.\label{f:ACCDF}}
\vspace{-0.1in}
\end{figure*}

In order to model the evolution of $B(t)$ and $\sN(\cB(t))$ we need to
understand the impact of SI recruitment policy on the degree distribution of the nodes still left to recruit, $\ocB(t)$.
Fig.~\ref{f:ACCDF} shows the empirical Complimentary Cumulative Distribution Function ($P[K \geq k]$), denoted CCDF in the plots, of nodes' degrees in $\cB(t)$ using SI over the Enron email network for $t \in \{$1000, 2000,4000,8000,16000$\}$ and against the CCDF of all the nodes in $V$.
The empirical CCDF is averaged over $1\,000$ runs.
Observe that even when $t$ is reasonably small, e.g., $t = 1,\!000$, the tail of the CCDF of $\ocB(t)$ is still significantly ``lighter'' than the tail of the CCDF of $V$. This is because large degree nodes are more likely to be recruited early to $\cB(t)$; and as $\ocB(t)$ is  depleted of large degree nodes, the tail of the degree distribution of $\ocB(t)$ gets ``lighter''.
We use this property when analyzing the cover performance of SI.

\subsection{Analysis of SI cover}
We start our analysis by characterizing the evolution of the cover size as a function of $t$.
Section~\ref{sec:changedist} shows that the number of remaining nodes in $\ocB(t)$ with degree $k=1,\ldots,N-1$ is a function of $t$ and $k$.
Hence, our analysis divides the recruited nodes in $\cB(t)$ into classes corresponding to different degrees.
In particular, using mean field approximations we characterize $b_k(t)$, the fraction of nodes of degree $k$ in $G$ that are recruited by time $t$, $k = 1,\ldots,N-1$.
As the number of nodes of degree $k$ in $\ocB(t)$ is given by $N p_k (1 -b_k(t))$,  the degree distribution of $\ocB(t)$ can be approximated by $\{C p_k (1 -b_k(t))\}_{k=1,\ldots,N-1}$, where $C$ is a normalizing constant.
We now characterize the connections between nodes of various degrees between $\cB(t)$ and $\ocB(t)$.
In the configuration model -- described in Sec.~\ref{sec:notation} -- the probability that a given node $u \in V$ of degree $k$ is connected to a randomly chosen node $X \in V$ of degree $k_X = h$ is
$$p_{kh}=1-\left(1-\frac{h}{2 M}\right)^k.$$
The probability that $X \in \cB(t)$ given $k_X = h$ is $P[X \in \cB(t) | k_X = h] = b_h(t)$, from which we
approximate the probability that an infected node $u$ of degree $k$ has an infected neighbor of degree $h$ by
$$P[X \in \cB(t) \, ,\, X \in \cN_a(u) | k_X = h, u \in \ocB(t), k_u=k] \approx b_h(t) p_{kh}.$$
Note that the above approximations assumes that $P[X \in \cB(t)]$ does not depend $u \in \ocB(t)$ (a reasonable approximation if $X$ is randomly chosen from a large population of $Np_h$ nodes in the graph).
The probability that no degree $h$ node connected to $u$ has been recruited at time $t$ can be approximated by $(1-b_h(t) p_{kh})^{N p_h }$.
If we condition on the event $u \in \ocB(t)$, then the probability that $u$  has at least one recruited neighbor is $\prod_{h}(1-b_h(t) p_{kh})^{N p_h }$.
Take notice that unconditioning the above on $P[u \in \ocB(t)] = (1-b_k(t))$ yields $\Upsilon_k(t)$, the probability that at time $t$ a randomly chosen node of $\ocB(t)$ of degree $k$ is in  $\cN(\cB(t))$, given by
$$
\Upsilon_k(t) = (1-b_k(t))\left(1-\prod_h(1-b_h(t) p_{kh})^{N p_h }\right).
$$
Finally, the expected number of observed nodes at time $t+1$, $\av{\sN(\cB(t+1))}$, is approximately
\begin{equation}
\label{e:gray_si}
\av{\sN(\cB(t+1))}\approx  \sum_k N p_k \Upsilon_k(t).
\end{equation}
Appendix~\ref{app:SIcontrast} contrasts~\eqref{e:gray_si} against previous works, e.g. Pastor-Satorras and Vespignani~\cite{vespignani02}.


Now we derive an equation for the dynamics of the number of sampled nodes $\cB(t)$.
When we sample a new edge $(u,v)$ with $u \in \cB(t)$ and $v \in \cN(\cB(t))$ from the frontier between $\cB(t)$ and $\cN(\cB(t))$, the probability that $v$ has degree $k$ is proportional to $k p_k (1-b_k(t))$. If we divide it by $p_k N$, we get the average increase in the fraction of sampled nodes of degree $k$, then:
\begin{equation}\label{eq:bk}
b_k(t+1)  =  b_k(t)+\frac{k (1-b_k(t))}{N \sum_h h p_h (1-b_h(t))}.
\end{equation}
It is easy to check that
\[
\av{B(t+1)}=\sum_k N p_k b_k(t+1)=\sum_k N p_k b_k(t) +1= \av{B(t)} +1,
\]
we guarantee that at every step a new node is recruited.


{

\begin{figure*}[t]
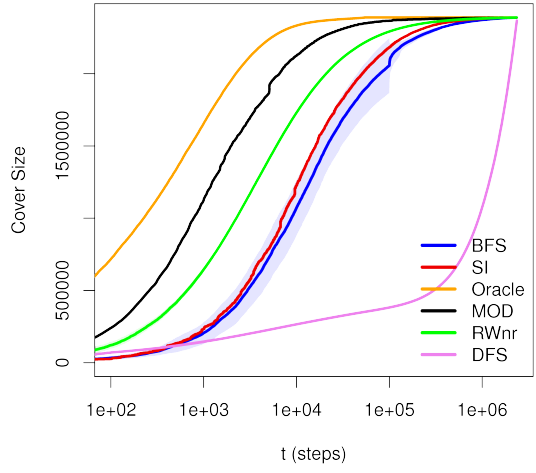
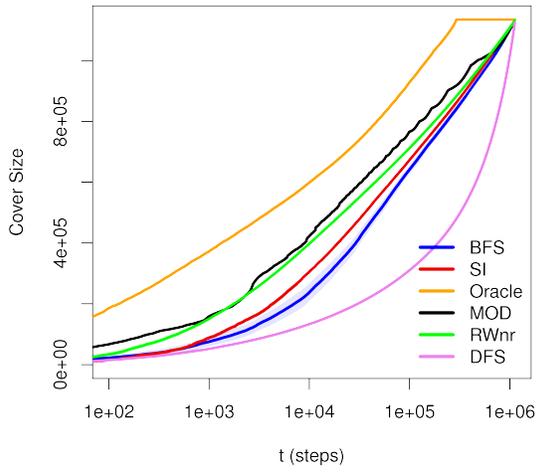
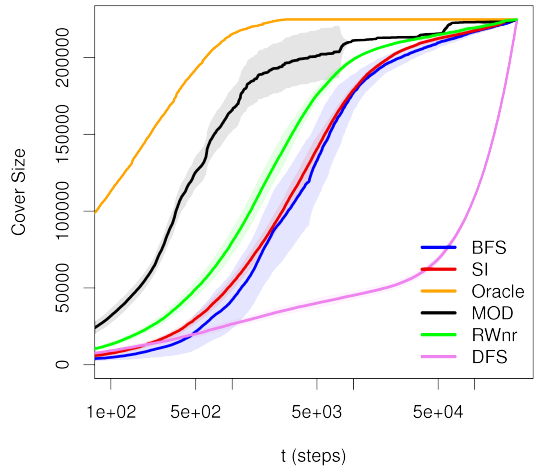
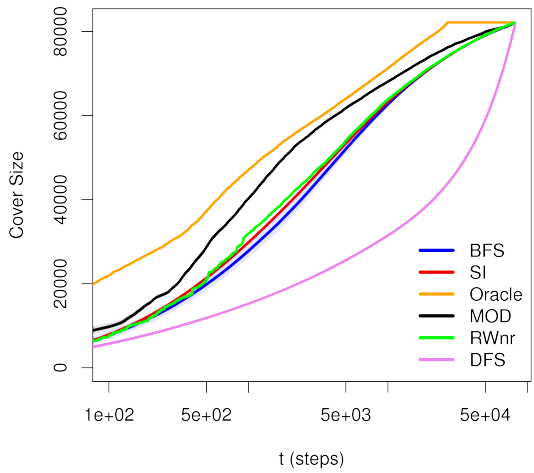
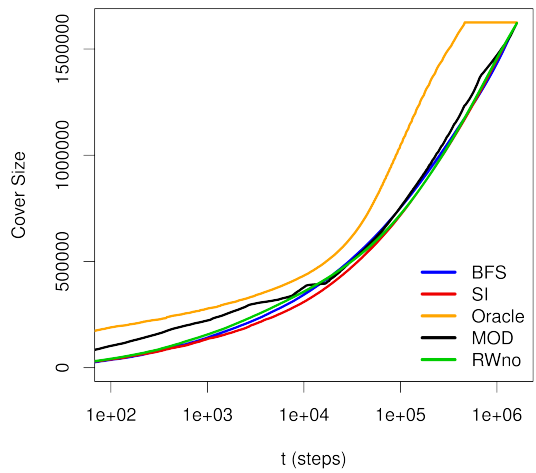

\vspace{-0.4in}
\centering
\subfloat[][Epinions Network]{
\includegraphics[width=3in,height=3in]{figs/effect_of_memory_Epinions_ORIG+MaxObs+GSI+RWno+BFS+DFS+MAXDEG_nodes_png.pdf}
}
\subfloat[][Wiki-talk Network]{
\includegraphics[width=3in,height=3in]{figs/effect_of_memory_Wikitalk_ORIG+MaxObs+GSI+RWno+BFS+DFS+MAXDEG_nodes_png.pdf}
}
\\
\vspace{-0.3in}
\subfloat[][Youtube Network]{
\includegraphics[width=3in,height=3in]{figs/effect_of_memory_Youtube_ORIG+MaxObs+GSI+RWno+BFS+DFS+MAXDEG_nostd_nodes_png.pdf}
}
\subfloat[][EmailEU Network]{
\includegraphics[width=3in,height=3in]{figs/effect_of_memory_emailEu_ORIG+MaxObs+GSI+RWno+BFS+DFS+MAXDEG_nodes_png.pdf}
}
\\
\vspace{-0.3in}
\subfloat[][Slashdot Network]{
\includegraphics[width=3in,height=3in]{figs/effect_of_memory_Slashdot_ORIG+MaxObs+GSI+RWno+BFS+DFS+MAXDEG_nodes_png.pdf}
}
\subfloat[][Flickr Network]{
\includegraphics[width=3in,height=3in]{figs/effect_of_memory_Flickr_ORIG+MOD+GSI+BFS+RWno+MAXDEG_nodes_png.pdf}
}
\caption{Empirical average cover size $\av{\oW(t)}$ of various social networks. Comparison between Oracle, RWnr, SI, BFS, DFS, and MOD algorithms. $x$-axis in log-scale. \label{f:social}}
\end{figure*}

We now contrast our mean field approximations against simulations.
Fig.~\ref{f:TheoSimul} shows $\av{\sN(\cB(t))}$ calculated according to~\eqref{e:gray_si} against the empirical value obtained from our simulations over two datasets, Enron and Gnutella.
We plot the results in log-log scale to facilitate the comparison in respect to the relative error.
Note that our approximation tracks the simulation results very well.
Fig.~\ref{f:GSIMOD} shows the SI cover size against the cover size of BFS and RWnr on the Enron email network.
Note that the SI cover size is larger than the BFS cover size but smaller than that of RWnr.
These results are consistent on all of the datasets analyzed, see Figs.~\ref{f:social}a-f, (albeit sometimes the differences are small).
The takeaway message from Fig.~\ref{f:GSIMOD} and all the results from the plots in Figs.~\ref{f:social}a-f is that there is much room for improvement.
And while RWnr clearly outperforms all other methods in the Enron dataset, this gain all but disappears in other datasets (Epinions, Slashdot, and Flickr).
In what follows we present an algorithm that consistently outperforms all of the algorithms studied so far.

\begin{figure*}[t]
\centering
\vspace{-0.4in}
\subfloat[][Enron Network]{
\includegraphics[width=3in,height=3in]{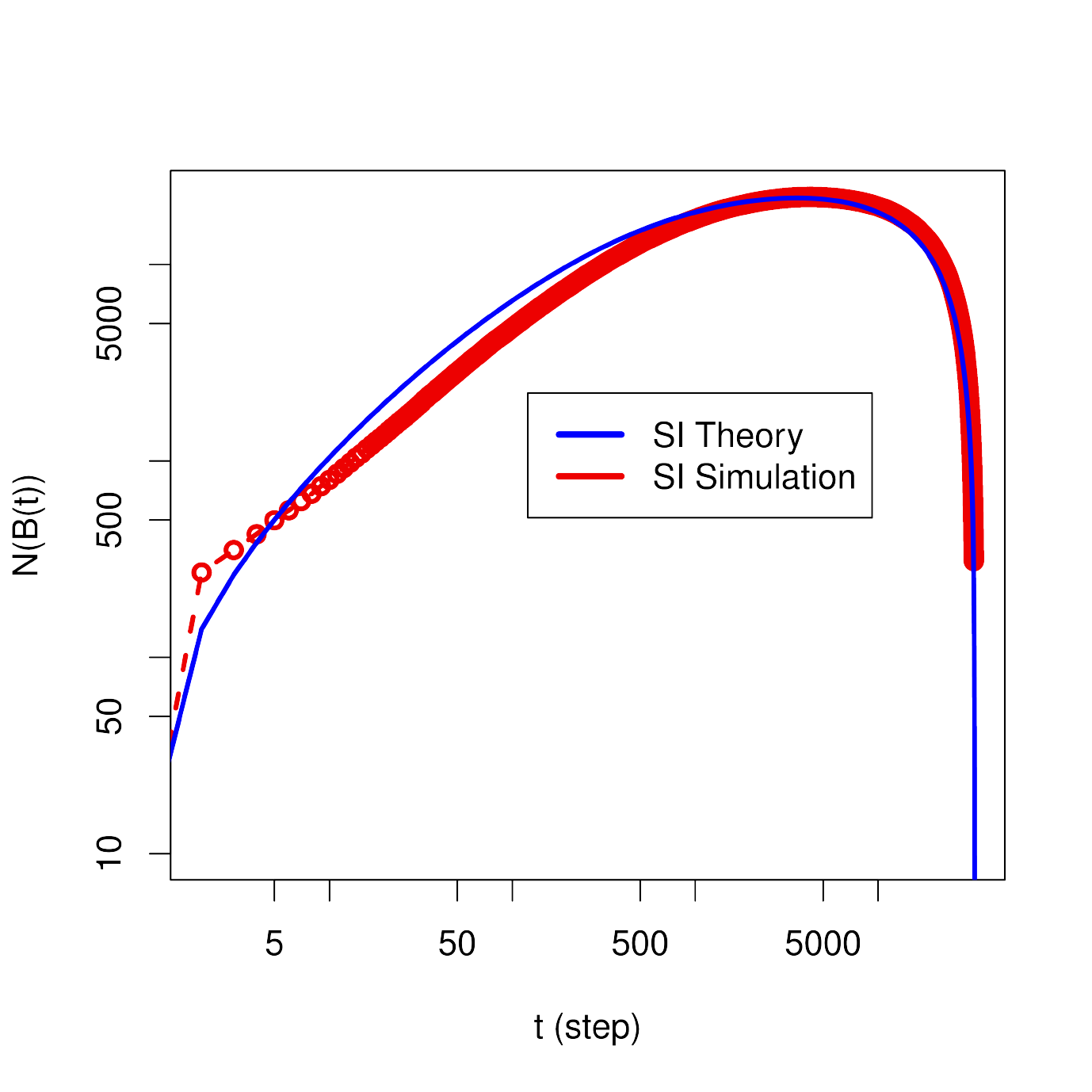}
}
\subfloat[][Gnutella Network]{
\includegraphics[width=3in,height=3in]{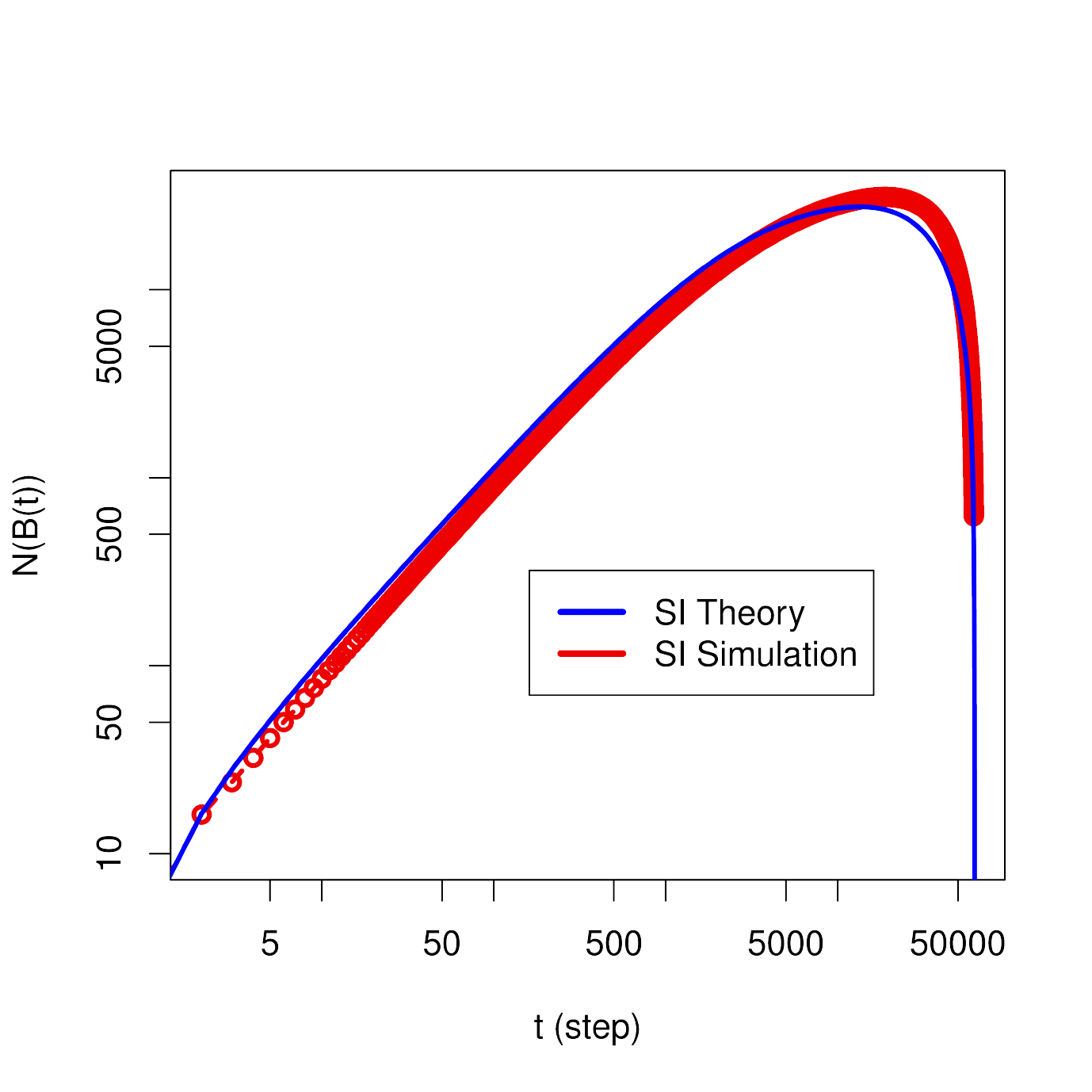}
}
\caption{\small%
Mean field approximation of $\av{\sN(B(t))}$ eq.~\eqref{e:gray_si} against the true average of $1,\!000$ simulations on Enron and Gnutella networks.\label{f:TheoSimul}}
\vspace{-0.1in}
\end{figure*}

\label{s:SI}


\section{Expected Excess Degree\\ Maximization Algorithm}
\label{sec:excess}

One lesson to take away from Guha and Khuller's Oracle is that knowing which node in $\cN(\cB(t))$ has the largest excess degree is crucial to achieving a good cover.
While the excess degree of nodes in $\cN(\cB(t))$ is not available to us, we may still be able to estimate them from the available information.
Let $d(v,t)$ be the {\em observed degree} of $v$ at step $t$.
Consider a large degree node $v \in \ocB(t)$. We expect $v$ to also have a large observed degree $d(v,t)$.
However, depending on the degree distribution of the nodes in $\cN(\cB(t))$, there could be a ``saturation point'', where the observed degree of $v$, $d(v,t)$, is so large that the excess degree $k_v - d(v,t)$ is small.
In this section we propose an algorithm, denoted {\bf Maximum Expected Excess Degree (MEED)}, that at step $t > 0$ finds a node in $\cN(\cB(t))$ that has approximately the largest expected excess degree of all nodes in $\cN(\cB(t))$.
In what follows we drop $t$ from our notation for the sake of conciseness.

Let $\av{k | d(v) }$ denote the expected degree of a node $v \in \cN(\cB(t))$ with observed degree $d(v)$.
Using $\av{k | d(v) }$ we propose the Maximum Expected $d$-Excess Degree (MEED) heuristic, which chooses the next recruited node $v^\star$ at step $t$ as to maximize the expected excess degree of the recruited node.
Another way to describe the MEED scheduler is through a partial order of the nodes in $\cN(\cB(t))$ in respect to their expected excess degree.
For two nodes $u,v \in \cN(\cB(t))$ we say $u \geq_k v$ iff  $\av{k - d(u) | d(u)} \geq \av{k - d(v) | d(v)}$.
Thus, at step $t$ the MEED algorithm recruits node $v^\star \in \cN(\cB(t))$ if $ v^\star \geq_k v$, $\forall v \in \cN(\cB(t))$.

\begin{figure*}[p]
\vspace{-0.1in}
\centering
\includegraphics[width=4.5in,height=3in]{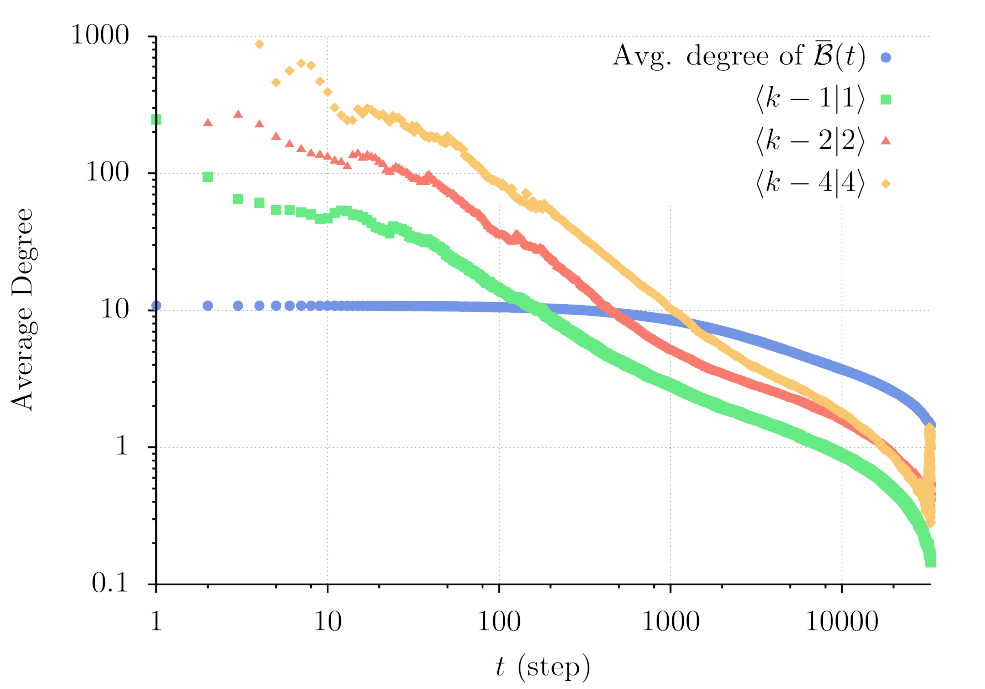}
\caption{\small%
{\bf (Enron Network)} Average $d$-excess degree of $\ocB(t)$ in SI as a function of step $t=1,\ldots,N-1$. Observe that $\av{k-4|4}$ is consistently larger for all values of $t$ than $\av{k-1|1}$ and $\av{k-2|2}$. Also note that this happens notwithstanding how little the average degree of $\ocB(t)$ varies over $t$ (blue dots).\label{f:d-excess}}
\vspace{-0.2in}
\end{figure*}


Next we obtain an approximation of $\av{k - d | d }$ using $\{p_k\}_{k=1,\ldots}$.
For now we assume $\{p_k\}_{k=1,\ldots}$ is given to us.
Later we show that for some important families of random networks, the node with maximum observed degree is also the node with the maximum expected excess degree.
Let $\zeta_k(t)$ be the probability that a random node in $\ocB(t)$ has degree $k$.
Note that $\zeta_k(0) = p_k $ as, by definition, the initial node in $\cB(0)$ is randomly sampled from $V$.
In general we have $\zeta_k(t) = C p_k (1 - b_k(t))$ (as stated in Sec.~\ref{sec:SI}), where $C$ is a normalization constant.
In what follows we omit $t$ for the sake of conciseness.

Let $\cN^{(d)}(\cB(t)) \subseteq \ocB(t)$ denote the set of nodes with at least $d$ recruited neighbors.
Note that $\cN^{(1)}(\cB(t)) = \cN(\cB(t))$ and that $\cN^{(0)}(\cB(t)) = \ocB(t)$.
Under the configuration model we can determine $\{\cN^{(d)}(\cB(t))\}_{d=1,\ldots}$ and $\cW(t)$ through the following process that dynamically assigns nodes from $\ocB(t)$ to these sets.
Let's assume $N \gg 1$ so we do not need to worry about self-loops.
Detach all nodes $v \in V$ from their neighbors such that node $v$ with degree $k_v$ has $k_v$ ``active stubs''.\footnote{This stub analogy is extensively used to describe configuration models~\cite{Newman}.}
Iteratively select an active stub in $\cB(t)$ to a random active stub in $V$.
Whenever an active stub of a node $u \in \cN^{(d)}(t)$, $d \in \{0,1,\ldots\}$, is selected, we add $v$ to $\cN^{(d+1)}(t)$ and mark both stubs of the edge ``inactive'', that is, we promote $u$ to $\cN^{(d+1)}(t)$ but reduce its active degree by one.
The following recursion describes the degree distribution $\{\zeta_k^{(d+1)}\}_{k=d+1,\ldots}$ of the nodes in $\cN^{(d+1)}(t)$ in terms of the degree distribution $\{\zeta_k^{(d)}\}_{k=d,\ldots}$ of nodes in $\cN^{(d)}(t)$
\vspace{-0.05in}
\begin{equation}\label{eq:gray}
   \zeta_k^{(d+1)} = \frac{(k-d) \zeta_{k}^{(d)}}{ \av{k}_{\zeta^{(d)}} - d  }, \quad  k \geq d,
\end{equation}
\vspace{-0.05in}
and $\zeta_k^{(d)} = 0$ for $k < d$,
where
$ \av{k}_{\zeta^{(d)}} \equiv \sum_{k \geq 0} k \zeta_{k}^{(d)}$ is the average degree of nodes with at least $d$ recruited (black) neighbors.

We now retrieve $\av{k |d} $ from $ \av{k}_{\zeta^{(d)}} $ using the fact that $\cN^{(0)} \supseteq \cN^{(1)} \supseteq \cdots$.
Let $N_d$ be the number of nodes in $\cN(\cB(t))$ with observed degree $d$.
Note that $N_1 = \sN(\cB(t))$.
For any two sets $A$, $A^\prime$, such that $A^\prime \subseteq A$, the following holds: $\vol(A - A^\prime) = \vol(A) - \vol(A^\prime)$, where $\vol(B)$ is the volume of the set $B$, i.e.~the sum of the all the degrees of the nodes in $B$.
Considering $A = \cN^{(d)}$ and $A^\prime = \cN^{(d+1)}$ it is easy to show that
$$
\av{k |d} = \Av{ \frac{N_{d}}{N_d - N_{d+1}}}  \av{k}_{\zeta^{(d)}} -  \Av{ \frac{N_{d+1}}{N_d - N_{d+1}}} \av{k }_{\zeta^{(d+1)}}  \, .
$$
We approximate the expectations of $\av{N_{d}/(N_d - N_{d+1})}$ and $\av{N_{d+1}/(N_d - N_{d+1})}$ using the observed values of $N_{d+1}$ and $N_{d}$.

Our calculations of  $\av{N_{d}/(N_d - N_{d+1})}$ and $\av{N_{d+1}/(N_d - N_{d+1})}$ should be used with caution as they do not consider the extra density of connections inside $\cB(t)$ created by the MEED recruitment process.
Taking this bias into account is not trivial and is the subject of future work.
However, in our MEED simulations we observe that $N_{d+1} \ll N_d$ for large $d$, and, under such scenario, it is reasonable to make the following simplification  
$
\av{k - d|d} \approx \av{k - d}_{\zeta^{(d)}} .
$

Unfortunately, obtaining $\av{k - d}_{\zeta^{(d)}}$ still requires knowing $\zeta_k^{(d)}$, $\forall k,d$, 
 which in turn requires knowing the degree distribution.
Note, however, that MEED simplifies to an algorithm that always selects the node $v^\star$ with the maximum observed degree in $\cN(\cB(t))$ if $\av{k - d(v^\star)|d(v^\star)} \geq \av{k - d|d}$, $d = 1,\ldots,d(v^\star)$.
We denote this simplified MEED heuristic {\bf Maximum Observed Degree (MOD)}.


We now see if we should expect to find the property $\av{k - d(v^\star)|d(v^\star)} \geq \av{k - d|d}$, $d = 1,\ldots,d(v^\star)$ in real social networks.
In what follows we show that, under certain conditions, two of the most relevant social network models -- power law and Er\"os-R\'enyi random networks -- have this desired property.
We later complement these findings with simulation results in the Enron network.
%
%
In Appendix~\ref{appx:dH} we show that
\begin{align}
    \av{k - d}_{\zeta^{(d)}}    & = \left. \frac{  \left( \frac{\partial ^{d+1}}{\partial z^{d+1}} H(z) \right)  } { \left( \frac{\partial ^{d} }{\partial z^{d}}H(z) \right) }\right|_{z = 1} \: , \quad d \geq 1 \, , \label{eq:kd}
\end{align}
where $H(z) = \sum_k z^k \zeta_k$ is the probability generating function (p.g.f.) of $\zeta_k$, and $\partial^d H(z)/\partial z^d$ is the $d$-th derivative of $H(z)$ with respect to $z$.
Next we use~\eqref{eq:kd} to obtain approximations of $\av{k - d}_{\zeta^{(d)}}$ for Erd\"os-R\'enyi and power law networks.

\subsection{Maximizing d-excess Degrees in \\ Erd\"os-R\'enyi Networks}
In an Erd\"os-R\'enyi (ER) graph $G(N,q)$ the degree distribution is Binomially distributed,
$p_k = \binom{N-1}{k} q^k (1 - q)^{N-1-k}$, $k=1,\ldots,N-1$.
We now consider the case where $t$ is small enough such that $\zeta_k(t) \approx p_k$.
Then, the probability generating function of $\zeta_k$ is $H(z) = (1 - q + q z)^{N-1}$.
In such scenario applying~\eqref{eq:kd} yields
\[
\av{k - d}_{\zeta^{(d)}} = \frac{ (N-1)! q^{d+1}/(N-d-2)! }{(N-1)! q^{d} /(N-d-1)!} =  (N-d-1) q  .
\]
As $N \gg 1$ we observe that the average $d$-excess degree can be approximated by the average degree of the network, $\av{k - d}_{\zeta^{(d)}} = \av{k}$, independent of $d$.
Thus, if $N_{d+1} \ll N_d$, $\forall d$, the scheduler can be degree agnostic and, thus, MOD approximates MEED.

\subsection{Maximizing d-excess Degrees in \\ Power Law Networks}
In a power law network we cannot ignore the effect of $t$ on $\zeta_k(t)$.
Hence, we consider $\zeta_k(t)$ to be power law distributed with an exponential cut-off.
The exponential cut-off approximates the behavior observed in Fig.~\ref{f:ACCDF}.
Moreover, in a variety of real world networks the degree distribution can be well approximated by power law distributions with exponential cut-offs~\cite{AmaralPL,NewmanPL}.
Let
$$
\zeta_k = \frac{k^{-\tau } C_t^k}{\Li_\tau(C_t)} \, , \quad \mbox{for } k\geq 1,
$$
where $C_t < 1$ is a parameter that depends on $\cB(t)$ and $t$ and the normalization factor $\Li_h(x) = \sum_{k=1}^\infty x^k / k^h$ is the $h$-th polylogarithm function of $x$.
The probability generating function of $\zeta_k$ assumes the form 
$
H(z) = \Li_\tau(z C_t)/\Li_\tau(C_t).
$
In Appendix~\ref{appx:dH} we use $H(z)$ to 
 show that:
\begin{itemize}
\item If $\tau = 1$, 
$
 \av{k - d}_{\zeta^{(d)}} \approx \frac{ C_t d}{1 - C_t}.
$
and thus the node with the largest observed degree should be recruited.
\item If $\tau = 2$,
$
 \av{k - d}_{\zeta^{(d)}}  \approx \Gamma_d   \, , \nonumber
$
where 
$
  \frac{d}{1+1/d}   < \Gamma_d \left( \frac{1 -C_t}{C_t} \right) <   d+1 \, ,
$
which implies that for $ \av{k - (d+a)}_{\zeta^{(d+a)}}  > \av{k - (d)}_{\zeta^{(d)}} $, $a = 2,3,\ldots$ and $d=1,\ldots$.
Thus, at step $t+1$ we should recruit the node $v \in \cN(\cB(t))$ with either the largest or second largest value $d(v)$.
We believe that these bounds can be improved to show that the node with the largest observed degree should be recruited.
An interesting observation is that $ \av{k - d}_{\zeta^{(d)}}$ increases with $C_t$ and diverges as $C_t \to 1$.

\item With $C_t \to 1$ and $\tau > 0$: 
The case where $C_t \to 1$ represents the case where $\{p_k\}$ is a pure power law distribution and $t$ has little impact on the degree distribution of $\ocB(t)$.
This case is only of theoretical interest as no real world network degree distribution can match an infinite support power law degree distribution.
In this scenario $ \av{k - d}_{\zeta^{(d)}} < \infty$, $\forall d \leq \ceil{\tau-1}$, and $ \av{k - \ceil{\tau}}_{\zeta^{(\ceil{\tau})}} \to \infty .$
This means that the node with observed degree degree at least $\ceil{\tau}$ should be recruited.
We believe that this result can be strengthen to show that $\av{k-d|d}$ monotonically increases with $d$.
\end{itemize}
Thus, we conclude that under the above networks MOD is a good approximation to MEED.

\subsection{Simulations, Excess Degree \& MOD}
The above analysis suggests that in some power law networks and in ER networks, $\av{k - d|d}$ increases with $d$. 
An important question is whether we observe this phenomenon in practice. 
Fig.~\ref{f:d-excess} shows estimates of $\av{k-d|d}$ for $d\in\{1,2,4\}$ obtained from simulating SI on the Enron network (these estimates are averaged over 1,000 runs). 
We choose SI to estimate $\av{k-d|d}$ instead of MEED as under MEED very few nodes with observed degrees greater than two remain in $\cN(\cB(t))$, $t > 0$, thus making the estimates unreliable.
In the figure we observe that for $t$ approximately in the range $\{1,\ldots,2/3N\}$ we have  $\av{k-1|1} < \av{k-2|2} < \av{k-4|4}$ and 
for $t$ approximately in the range $\{2/3N+1,\ldots,N-1\}$ we have $\av{k-1|1} < \av{k-2|2} \approx \av{k-4|4}$.
Hence we expect that MOD is a good candidate to approximate MEED on Enron network.

Fig.~\ref{f:GSIMOD} and Figs.~\ref{f:social}a-f show that the MOD heuristic outperforms, sometimes significantly, all previous lookahead one algorithms on all social network datasets.
We, however, still notice a significant gap between the Oracle and MOD, which we believe can be reduced using side information 
to improve the estimation of $ \av{k - d}_{\zeta^{(d)}}$.

%
\section{Summary, Conclusions \& \\ Related Work}\label{sec:related}
We have considered the problem of providing an online algorithm that, by recruiting nodes through their neighbors, greedily maximizes the network cover of an online social network.
In our setting the network topology was unknown and the only topological information available came from the identity of the neighbors of already recruited nodes.

In this scenario, we have evaluated the efficacy of existing network sampling algorithms (BFS, DFS, RW) and proposed a novel algorithm, Maximum Expected Excess Degree (MEED), inspired by the greedy approximation to the minimum connected dominating set of Guha and Khuller~\cite{Guha1998}, which uses two-hop lookaheads (and, thus, denoted ``Oracle'' in this work) to recruit at every step the node with the largest excess degree.
The MEED heuristic seeks to maximize the expected excess degree of nodes with the help of degree distribution side information (if available).
In the absence of degree distribution information, we have shown that on random power law and Erd\"os-R\'enyi networks  MEED can be approximated by MOD (Maximum Observed Degree), a greedy heuristic that at every step recruits the node with the largest observed degree.
We have shown through extensive simulations on real world social network datasets that MOD outperforms all other algorithms, often quite significantly.

We have also provided theoretical analysis of RWs (with and without replacement) and of an algorithm inspired by the Susceptible-Infected epidemic model, which we denoted SI.
Our theoretical analysis, to the best of our knowledge, stands as a contribution on its own. We expect that our formulas can aid practitioners in predicting the cover sizes of these algorithms when degree distribution side information is available. 

Finally, we have uncovered a puzzling previously unknown fact about DFS: DFS performs remarkably poorly  on social networks.
In fact, DFS seems to avoid recruiting nodes with large excess degrees.
We have argued that this is due to its tendency to keep large degree nodes at the bottom of the its recruitment queue.
We note in passing that this property of DFS may find applications in undercover military operations where 
one seeks to recruit target individuals with the minimum exposure (number of connections) to unrecruited targets.

\paragraph{Related Work}
The connections between our work and the literature on MCDS were already presented in Sec.~\ref{sec:opt}.
In this section we review the remaining related literature.
The work most related to ours is Maiya and Berger-Wolf~\cite{MaiyaKDD}.
Maiya and Berger-Wolf presents a simulation study of the cover sizes (among other metrics) of different algorithms, including BFS, DFS, Oracle (which they denote {\em Expansion Sampling}), and MOD (which they denote {\em Sample Edge Count}).
Their work considers social (e.g., Enron email) and technological (e.g., Amazon product co-purchase) networks.
Surprisingly, their conclusions are remarkably different than ours, arguing that DFS outperforms BFS, RWs, and, most importantly, MOD.

Maiya and Berger-Wolf~\cite{MaiyaKDD} shows the performance of these algorithms, Figs.~4(e) and~4(f) in their work, on the (non-social) HepTh and Amazon networks (see Sec.~\ref{sec:datasets} for a brief description of these datasets).
In order to understand the discrepancy of our conclusions, here we also include simulations of these same two datasets.
We show our results in Figs.~\ref{f:HP} and~\ref{f:AMZ}.
Note the remarkable difference to our social network results in Figs.~\ref{f:Enron}a-b and~Figs.\ref{f:social}a-f.
Indeed, DFS experiences a great improvement in performance while MOD worsens dramatically. 

To test whether the sudden improvement of DFS and decline of MOD can be mostly attributed to
the more structured nature of these two networks in respect to our social networks, we artificially add randomness to these network by randomly rewiring all endpoints (while making sure nodes form a single connected component).
Figs.~\ref{f:RHP} and~\ref{f:RAMZ} provide the results over these randomized networks.
The results are clear, adding randomness makes MOD incontestably superior to all other algorithms (in HepTh it even matches the performance of Oracle) and DFS is again noticeably inferior.
It remains an open question whether, in respect to the network cover problem, most social networks are more similar to ``random networks'' or more similar to ``structured networks''. Our datasets and simulation results suggest the former.

\begin{figure*}[t]
\vspace{-0.4in}
\centering
\subfloat[][HepTh\label{f:HP}]{
\includegraphics[width=3in,height=3in]{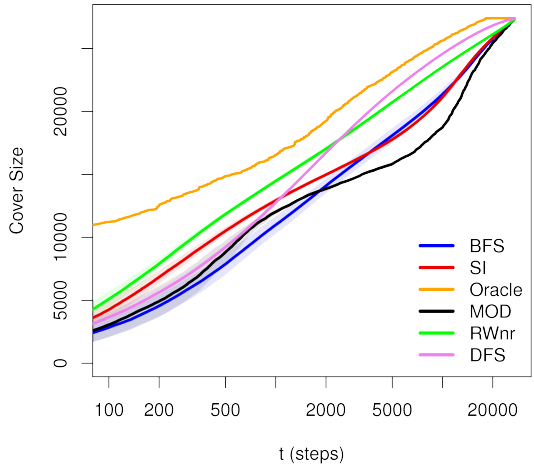}
}
\subfloat[][Amazon\label{f:AMZ}]{
\includegraphics[width=3in,height=3in]{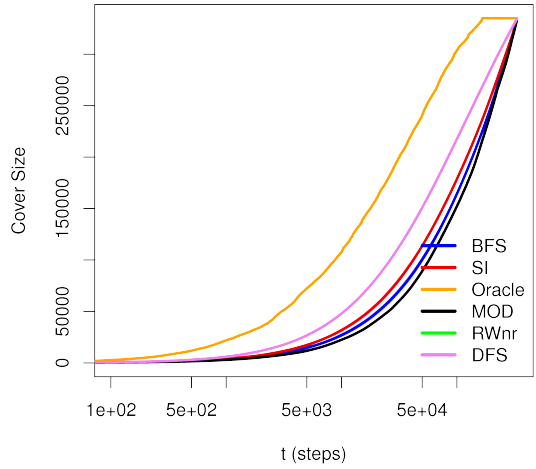}
}
\\
\vspace{-0.3in}
\subfloat[][Randomized HepTh\label{f:RHP}]{
\includegraphics[width=3in,height=3in]{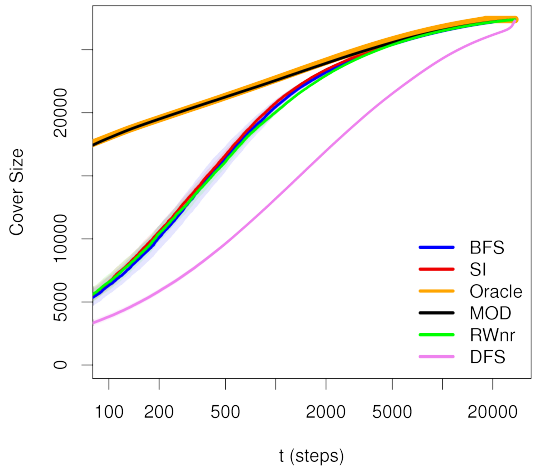}
}
\subfloat[][Randomized Amazon\label{f:RAMZ}]{
\includegraphics[width=3in,height=3in]{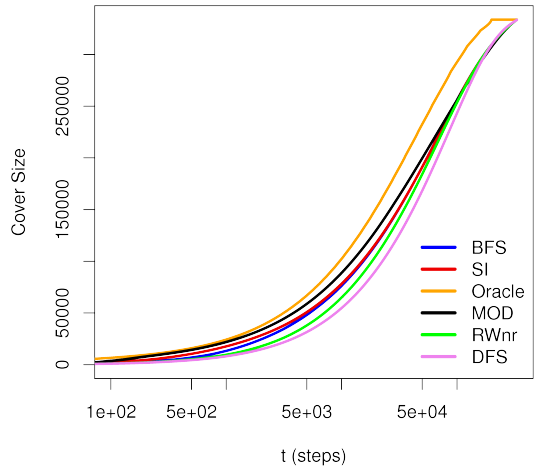}
}
\caption{\small%
The two datasets in which DFS outperforms MOD. Comparison between the empirical average cover size $\av{\oW(t)}$ of Oracle, RWnr, SI, BFS, DFS, and MOD algorithms. Figs.~\ref{f:HP} and~\ref{f:AMZ} show the results on the original networks and Figs.~\ref{f:RHP} and~\ref{f:RAMZ} show the results in their randomized counterparts. Note that when randomized, we see similar results seen in the social networks. Thus, the good performance of DFS and poor performance of MOD are due to the peculiar network topology of these graphs. $x$-axis in log-scale. \label{f:nonsociall}}
\vspace{-0.1in}
\end{figure*}

In an earlier preliminary work (Lim et al.~\cite{Lim11}) we proposed SI under the name {\em Randomized Expansion Sampling} (RXS) used to find the most central nodes in a network. However, in Lim et al.\ we did not analyze SI cover size.
Random walks with lookahead have been the subject of a number of works.
Cooper and Frieze~\cite{Cooper-RW-Lookahead-10} studied the cover time of RW.
Mihail et al.~\cite{MST06} shows that a RW with one-hop lookahead finds the majority of nodes in sublinear time in an infinite configuration model with
heavy tailed power law degree distribution. 
Our analysis, however, shows that the cover rate is at most linear for any value of $t$. 
Adamic et al.~\cite{Adamic} proposed a RW with two-hop lookahead and analyzed its cover time on an infinite network.
The fast cover of RWs has been used in the context of decentralized search, e.g., when searching for content on unstructured P2P networks (see~\cite{KSearch,Duff,tsoumakos2006analysis,LCCLS02,Ioannidis09} and references therein).

A closely related problem is the {\em influence maximization} problem.
The influence maximization problem considers that each recruited individual invites its neighbors who can be recruited with some probability. The purpose is to select a set ``influential'' individuals, in order to cause a cascade of recruitments in the network. Network topology is generally assumed to be known~\cite{Kempe}.
This problem was first proposed by Domingos and Richardson~\cite{Domingos01} (please refer to the review in Kleinberg~\cite{Kviral} for other references).
Directly related to online recruitment, Hartline et al.~\cite{HMS08} and Bhattacharya et al.\cite{BKC12} analyze the (paid) recruitment of consumers of a product under the assumption of perfect network knowledge.

\appendix

\section{RW \& Independent Edge Sampling Approximation}
\label{appx:RW}
Let us assume that the random walk starts from a stationary distribution ($\pi_v=k_v/(2M)$).
This is not too restrictive assumption, since in the configuration model after
the first step the random walk is asymptotically -- with respect to the network
size -- in the stationary regime. And in the general case, we can add auxiliary
uniform jumps, which has been shown~\cite{ART10} to significantly reduce mixing time.

We now show that if the random walk starts in steady state, that is $q_v=\pi_v$, the
expression using the complete topological information~\eqref{eq:EWkRW} can be approximated by expression~\eqref{eq:EWkRWsteady}.
It is enough to consider just one term in (\ref{eq:EWkRW}) corresponding to
one node. Without loss of generality, we take $v=1$ and assume that nodes $u=2,...,k_1+1$ are
neighbors of node 1. Partition the stationary distribution $\pi$ as $[\pi_1 \ \pi_2]$, where
$\pi_1$ corresponds to node 1 and all its neighbors and $\pi_2$ corresponds to all the other nodes.
Then, we need to evaluate
$$
\pi \ {}_{{\cal N}(v)}P = \pi_2 P_{22}^k \ones.
$$
Let us demonstrate that $\pi_2$ properly normed is close to $\tilde\pi_2$, the quasi-stationary
distribution of the substochastic matrix $P_{22}$:
$$
\pi_2 P_{22} = \lambda \pi_2.
$$
Even though $P_{22}$ is substochastic, if the graph is large enough, $P_{22}$ will be close
enough to a stochastic matrix so that we can apply
perturbation theory. Let us consider the following perturbation equation
\begin{equation}\label{eq:perteq}
(\pi_2^{(0)}+\eps \pi_2^{(1)}+...)(S_{22}-\eps D) = (1-\eps \lambda^{(1)}+...)(\pi_2^{(0)}+\eps \pi_2^{(1)}+...),
\end{equation}
where $S_{22}$ is the stochastic complement (it describes the transitions of the censored Markov
chain) and
$$
\eps D = P_{21} [I-P_{11}]^{-1} P_{12},
$$
with $\eps$ as some scaling parameter. Equating terms in (\ref{eq:perteq}) with $\eps^0$, we
obtain
$$
\pi_2^{(0)} S_{22} = \pi_2^{(0)},
$$
from which it follows that $\pi_2^{(0)} = c \pi_2$. Then, equating terms in (\ref{eq:perteq})
with $\eps^1$, we obtain
$$
\pi_2^{(1)} S_{22} - \pi_2^{(0)} D = \pi_2^{(1)} - \lambda^{(1)} \pi_2^{(0)}.
$$
Multiplication the above equation by $\ones$ from the right yields
$$
\lambda^{(1)} = \frac{1}{\pi_2 \ones} \pi_2 D \ones,
$$
and consequently,
$$
\lambda \approx 1 - \eps \lambda^{(1)}
= 1 - \frac{1}{\pi_2 \ones} \pi_2 P_{21} [I-P_{11}]^{-1} P_{12} \ones.
$$
Next, from the defining equations for the stationary distribution, we have
$$
\pi_2 P_{21} [I-P_{11}]^{-1} = \pi_1,
$$
which leads to
$$
\lambda \approx 1 - \frac{1}{\pi_2 \ones} \pi_1 P_{12} \ones.
$$
First, we note that $\pi_2 \ones \approx 1$. And second, we note that $\pi_1 P_{12} \ones$
is the number of links from the neighborhood set ${\cal N}(1)$ to all the other links
divided by the total number of links. Since the number of links to the outside of the
neighborhood set is much large than the number of links inside the set,
$\lambda \approx (1-\alpha_1)$ and expression~\eqref{eq:EWkRW} can be approximated
by~\eqref{eq:EWkRWsteady}.

\section{RW without replacement} \label{apps:RWnr}
Here we provide a brief description of the steps used to derive the approximation
in equations~\eqref{eq:EWkRWnrreplace}.

As in the case of the analysis of RW with replacement, we assume that we sample nodes
i.i.d. fashion according to the stationary distribution.
Recall that $Z(t)$ denotes the number of edges not yet discovered at time $t$. Then,
given the history $\{Z(i), i=0,1,...,t-1\}$ with $Z(0)=M$, the probability that the node $v$
is uncovered at time $t$ is given by
\begin{align*}
&P[v \in {\cal W}(t)|\{Z(i), i=1,...,t-1\}]\\
& = \prod_{i=0}^{t-1}\left(1-\frac{1}{2 Z(i)}
\left(k_v + \sum_{j \in \cN_a(v)} k_j\right)\right) \, ,
\end{align*}
Similarly, given the history $\{Z(t), t=0,1,...,t-1\}$, we can calculate the probability
that the link $(u,v)$ is yet uncovered at time step $t$
$$
P[\mbox{link $(u,v)$ is uncovered at $t$}|\{Z(i), i=1,...,t-1\}]=
$$
$$
\prod_{i=0}^{t-1} \left(1-\frac{k_u+k_v}{2Z(i)}\right).
$$
Viewing the link coverage problem as a subset coupon collector problem, we obtain
$$
\av{Z(t)|\{Z(i), i=1,...,t-1\}} = \sum_{(u,v) \in E} \:
\prod_{i=0}^{t-1} \left(1-\frac{k_u+k_v}{2Z(i)}\right).
$$
If we use the above equation recursively starting at $Z(0)=M$ and using the approximate
value of $\av{Z(t)}$, we obtain the second equation in the approximate formulas~\eqref{eq:EWkRWnrreplace}.

\section{BFS on 2D and 3D Grids}
\label{appx:grids}
We first observe that BFS discovers the graph by considering progressively large \emph{spheres}. At step $k$ the boundary of the set of discovered nodes is made by all nodes $k$ hops away from the first node. When nodes are embedded in a metric space, the sphere has the property to be the solid with the smallest surface for a given volume. This property justifies intuitively why BFS discovers slowly new nodes.

We first consider an infinite 2D grid. For the 2D grid there are $4 t$ nodes at the boundary at step $t$, then every node on the boundary contributes on the average to add $1+1/t$ new nodes when it is visited, even if each of them has degree $4$ and at least $2$ neighbors not discovered.

For a 3D grid the boundary has size $2+4 t^2$, then every nodes contributes to discover $(1+2(t+1)^2)/(1+2t^2)$. Here a node on the boundary has $6$ neighbors and at least $3$ not discovered at the begin of the step, but still its average contribution converges to $1$ as $t$ diverges.

The reason for this slow increase is due to the fact that 1) nodes at the boundary have many connections (roughly half of them, i.e. $\av{k}/2$) to the interior of the sphere, i.e. to already explored or discovered nodes, but also 2) the outgoing connections point to the same nodes. In fact almost every node that is going to be discovered will be discovered through $\av{k}/2$ connections (those pointing towards the sphere). Then, as the sphere becomes larger and the border effects negligible, every node only contributes by discovering one new node.

\section{SI Algorithm vs.\ SI Epidemic}
\label{app:SIcontrast}
The literature on SI epidemic models is so vast and rich that we dedicate a section of our appendix to contrast our results against previous works.
We also provide limited commentary on approximations of our equations.
We first note that the previous literature is often interested in a continuous-time version of our SI model.
Recall that in our scenario $t$ is the number of recruited nodes, while the related literature considers $t$ as time~\cite{vespignani02,vespignani02_2}, where at time $t$ the average number of ``infected nodes'' may be smaller or larger than $t$.

However, it is an interesting exercise to try to connect the framework provided by Pastor-Satorras and Vespignani~\cite{vespignani02} with our approach.\footnote{\small%
The authors would like to acknowledge N.\ Perra and A.\ Baronchelli for helpful discussions on this topic.}
Consider a {\bf continuous-time SI epidemic} on $G$.
In an SI epidemic an infected node can be thought play the role of a recruited node and susceptible nodes play the role of non-recruited nodes.
Let $\lambda$ be the per-unit time infection rate, that is, an infected node contaminates (recruits) a susceptible node during a time interval $\Delta \to 0$ with probability $\lambda \Delta t$, regardless of the state of the infection.
Let $t^\prime$ denote the {\bf wall clock time} of the SI epidemic process and $\rho_k(t^\prime)$ be the probability that a node with $k$ links is infected.
Then~\cite{vespignani02},
\begin{equation}\label{eq:vesp}
\frac{d \rho_k(t^\prime)}{dt} = \lambda k (1-\rho_k(t^\prime))\Theta(\rho(t^\prime)), \quad \forall k,
\end{equation}
where $\rho(t^\prime) = (\rho_1(t^\prime),\ldots,\rho_N(t^\prime))$ and
\begin{equation}\label{eq:Theta}
\Theta(\rho(t^\prime)) = \av{k}^{-1} \sum_k k p_k \rho_k(t^\prime)
\end{equation}
is the probability that any given link points to an infected node.
Note that our $t$ is the number of infected nodes, and thus,
\begin{equation}\label{eq:tprime}
\sum_k N p_k  \rho_k(t^\prime)  = t.
\end{equation}
Analytically adding the constraint~\eqref{eq:vesp} into the set of equations in~\eqref{eq:tprime} is not trivial, but it can be done numerically.
Let $\rho^\prime(t)$ be the solution of~\eqref{eq:tprime} with the added constraint~\eqref{eq:vesp}.
Still, even with $\rho^\prime(t)$ it is unclear how the cover size can be derived from~\eqref{eq:Theta}.
The main difficulty is the mapping between wall-clock time and number of recruited nodes. Our formulation in Section~\ref{sec:SI} solves this problem by avoiding formulation the problem in terms of wall-clock time.

%
%
%
%
%
\section{The expected {\large $d$}-excess degree from the p.g.f.\ of {\large $\zeta_k$}}
\label{appx:dH}
Our analysis of $\av{k-d|d}$ begins by breaking down $\av{k - d}_{\zeta^{(d)}}$ into the derivatives of the generating function of $\zeta_k$:
\begin{align}
\av{k - d}_{\zeta^{(d)}} &= \sum_{k=d}^\infty (k-d) \zeta_{k}^{(d)} \nonumber \\
  				 & = \sum_{h=0}^\infty  \frac{ h (h+1) \zeta_{h+d}^{(d-1)} } {   \av{k}_{\zeta^{(d-1)}} - (d-1) } \nonumber \\
  				 & = \frac{ \sum_{h=0}^\infty h (h+1) (h+2) \zeta_{h+d}^{(d-2)}} {  \prod_{i=d-2}^{d-1}  (\av{k}_{\zeta^{(i)}} - i) } \nonumber \\
                                 & = \frac{ \sum_{h=0}^\infty h (h+1) \cdots (h+d) \zeta_{h+d} } {  \prod_{i=0}^{d-1}  (\av{k}_{\zeta^{(i)}} - i) } \nonumber \\
                                 & = \frac{ 1} {  \prod_{i=0}^{d-1}  (\av{k}_{\zeta^{(i)}} - i) } \left. \frac{\partial ^{d+1} H(z)}{\partial z^{d+1}} \right|_{z = 1} \label{eq:pgf} ,
\end{align}
where $H(z) = \sum_k z^k \zeta_k$ is the probability generating function of $\zeta_k$.
The first several equalities are a consequence of successive applications of~\eqref{eq:gray}.
The last equality comes from the applying the $d+1$-th derivative to $H(z)$.
Multiplying both sides of~\eqref{eq:pgf}  by $ \prod_{i=0}^{d-1}  (\av{k}_{\zeta^{(i)}} - i) $ yields
\begin{align}
\prod_{i=0}^{d}  (\av{k}_{\zeta^{(i)}} - i)    & = \left. \frac{\partial ^{d+1} H(z)}{\partial z^{d+1}} \right|_{z = 1} \label{eq:kd_rec} ,
\end{align}
where $ \av{k}_{\zeta^{(0)}} = \sum_k k p_k$.
Substituting~\eqref{eq:kd_rec} into the denominator of the l.h.s.\ of~\eqref{eq:pgf} yields
\begin{align}
    \av{k - d}_{\zeta^{(d)}}    & = \left. \frac{  \left( \frac{\partial ^{d+1} H(z)}{\partial z^{d+1}} \right)  } { \left( \frac{\partial ^{d} H(z)}{\partial z^{d}} \right) }\right|_{z = 1} \: , \quad d \geq 1. \label{eq:kd}
\end{align}
An important property of the polylogarithm function is that for any constant $C > 0$,
\[
 \frac{\partial \Li_\tau(C z)}{\partial z} = \frac{\Li_{\tau-1}(C  z)}{z}.
\]
IN what follows we consider three special cases: $\tau=1$, $\tau=2$, and $C_t \to 1$.
\\

\noindent
{\bf Case $\tau = 1$}: Let's first consider the case $\tau = 1$, which implies that $\zeta_k$ is very heavy tailed.
We use the fact that $\Li_{1}(a) = - \log (1 - a)$.
With $\tau=1$ we have
\[
 \frac{\partial H(z)}{\partial z} = -\frac{1}{ \log (1 - C_t)} \frac{\partial \Li_1(C_t  z)}{\partial z} = -\frac{1}{\log(1-C_t) (1 - C_t z)}
\]
and thus,
\[
 \frac{\partial^{d} H(z)}{\partial z^d} = -\frac{1}{\log (1 - C_t)} \frac{(d-1)!C_t^d}{(1-C_t)^d},
\]
which yields
\[
 \av{k - d}_{\zeta^{(d)}} \approx \frac{ C_t d}{1 - C_t},
\]
showing that the $d$-excess degree grows linearly with $d$ when $\tau = 1$.
Then, we conclude that recruiting the node with the largest observed degree from $\cN(\cB(t))$ maximizes the expected cover increase.
\\

\noindent
{\bf Case $\tau = 2$}: Another case of interest is $\tau = 2$.
In this case note that
\begin{align*}
 \av{k - d}_{\zeta^{(d)}} & \approx \left. \frac{\frac{ \partial^d }{\partial z^d} \left( - \log(1-C_t z)/z \right)}{\frac{ \partial^{d-1} }{\partial z^{d-1}} \left( - \log(1-C_t z)/z \right)} \right|_{z=1} \: .
\end{align*}
Using the Taylor series expansion of
$$
- \log(1-C_t z)/z = \sum_{h \geq 1} \frac{C_t^h z^{h-1}}{h}
$$
yields
\[
 \av{k - d}_{\zeta^{(d)}}  \approx \Gamma_d   \, , \nonumber
\]
where
\begin{equation}
\begin{aligned}
\Gamma_d & =\frac{\sum_{h\geq d+2} \frac{C_t^h (h-1)!}{h (h-d-2)!}}{\sum_{h\geq d+1} \frac{C_t^h (h-1)!}{h (h-d-1)!}} \\
& =  \frac{ \sum_{m\geq 0} \frac{C_t^{m+1} (m+d+1)! }{(m+d+2) m!}}{\sum_{m\geq 0} \frac{C_t^m (m+d)!}{(m+d+1) m!}} . \label{eq:dav}
\end{aligned}
\end{equation}
We now obtain upper and lower bounds of eq.~\eqref{eq:dav} for any $d=1,2,\ldots$.
To derive a lower bound note that
\[
  \frac{m+d+2}{(1+1/d)(m+d+1)} < 1
\]
and that $(m+d)/(m+d+1) < 1$, for $m=0,1,\ldots$.
Using the above we decrease the numerator and increase the denominator of~\eqref{eq:dav} obtaining the lower bound:
\[
\Gamma_d >
\frac{ \sum_{m\geq 0} \frac{C_t^{m+1} (m+d+1)! }{ (1+1/d) (m+d+1) m!}}{\sum_{m\geq 0} \frac{C_t^m (m+d)!}{(m+d) m!}} = \frac{C_t d}{(1+1/d) (1 -C_t)}.
\]
Similarly, an upper bound of~\eqref{eq:dav} can be obtained from applying the inequalities
\[
  \frac{m+d+1}{(1+1/d)(m+d)} \leq 1
\]
and $(m+d+1)/(m+d+2) < 1$, for $m=0,1,\ldots$ into~\eqref{eq:dav}
\[
\Gamma_d <
\frac{ \sum_{m\geq 0} \frac{C_t^{m+1} (m+d+1)! }{ (m+d+1) m!}}{\sum_{m\geq 0} \frac{C_t^m (m+d)!}{ (1+1/d) (m+d) m!}} = \frac{(d+1) C_t}{ (1 -C_t)}.
\]
Thus,
\[
  \frac{d}{1+1/d}   < \Gamma_d \left( \frac{1 -C_t}{C_t} \right) <   d+1
\]
which implies that for $\av{k - (d+a)| d+a} > \av{k - d}_{\zeta^{(d)}}$, for all $a \geq 2$ and $d=1,\ldots$.
This means that at step $t+1$ we should recruit the node $v \in \cN(\cB(t))$ with either the largest value $d(v)$ of all nodes in $\cN(\cB(t))$ or the second largest value.
We hypothesize that improving the above bounds will reveal that node $v$ should have the largest value of $d(v)$.
Thus, if the above hypothesis holds, recruiting the node with the largest observed degree from $\cN(\cB(t))$ maximizes the expected cover increase.
An interesting observation is that $ \av{k - d}_{\zeta^{(d)}}$ increases with $C_t$ and diverges as $C_t \to 1$.
We now explore the case $C_t \to 1$.
\\

\noindent
{\bf Case $C_t = 1$}: The case when $C_t = 1$ represents the case when $\{p_k\}$ is a pure power law distribution --  that is, $\zeta_k = k^{-\tau}/\zeta(\tau)$, where $\zeta(\tau)$ is the Riemann zeta function with parameter $\tau$ (note that $\Li_{\tau}(1) = \zeta(\tau)$) -- and $t$ has little impact on the degree distribution of $\ocB(t)$.
This case requires assuming $N \to \infty$ and $t = o(N)$ and thus
it is just of theoretical interest as no real world network is an infinite power law network.
Because $\zeta(a) \to \infty$ for $a \leq 1$ and $\zeta(a) < \infty$ for $a > 1$, we observe that
\[
\frac{\partial ^{\ceil{\tau-1}} H(z)}{\partial z^{\ceil{\tau-1}}}    < \infty
\]
and
\[
\frac{\partial ^{\ceil{\tau}} H(z)}{\partial z^{\ceil{\tau}}}    \to \infty  \, .
\]
Thus, $ \av{k - d|d } $ diverges for $d=\ceil{\tau}$ and converges for $d < \ceil{\tau}$.
This implies that the expected $d$-excess degree of a node with $d=\ceil{\tau}$ recruited neighbors is infinite.
Due to the appearance of subtractions between two infinite quantities we were unable to verify whether
$\av{k - d|d } \to \infty$ for all $d > \ceil{\tau}$ or, more importantly, whether or not $ \av{k - d-1|d+1 } / \av{k - d|d } > 1$ holds for all $d > \ceil{\tau}$.
We, however, hypothesize that $\av{k - d|d } \to \infty$ for all $d \geq \ceil{\tau}$.
And again we see that recruiting the node with the largest observed degree from $\cN(\cB(t))$ maximizes the expected cover increase.

\balance

\bibliographystyle{plain}


\end{document}